\documentclass[twocolumn]{aastex6}

\usepackage{graphicx}
\usepackage{amsmath}
\usepackage{icomma}
\usepackage{psfig}
\usepackage{color}

\def\nm{\textrm{nm}}
\def\um{\mu\textrm{m}}

\def\pc{\textrm{pc}}
\def\kpc{\textrm{kpc}}

\def\yr{\textrm{yr}}

\def\Myr{\textrm{Myr}}
\def\Gyr{\textrm{Gyr}}

\def\Jy{\textrm{Jy}}

\def\Kelv{\textrm{K}}

\def\kgm2{\textrm{kg\ m}^{-2}}
\def\mgm2{\textrm{mg\ m}^{-2}}
\def\ugm2{\mu\textrm{g\ m}^{-2}}

\def\Msun{\textrm{M}_{\odot}}
\def\Lsun{\textrm{L}_{\odot}}
\def\Zsun{\textrm{Z}_{\odot}}
\def\LMsun{\Lsun/\Msun}

\def\GHAT{\^G}
\def\icarus{Icarus}

\def\ga{\gtrsim}
\def\la{\lesssim}

\def\endash{\text{--}}

\newcommand{\mean}[1]{\ensuremath{\langle #1 \rangle}}

\newcommand{\vIIChange}[1]{#1}
\newcommand{\vIIIChange}[1]{#1}

\shorttitle{Partially Cloaked Galaxies}
\shortauthors{Lacki}

\begin{document}

\title{Sunscreen: Photometric Signatures of Galaxies Partially Cloaked in Dyson Spheres}
\author{Brian C. Lacki$^{1}$}
\affiliation{Breakthrough Listen, Astronomy Department, University of California, Berkeley, CA, USA}
\email{astrobrianlacki@gmail.com}

\begin{abstract}
The Search for Extraterrestrial Intelligence has so far come up negative for Kardashev Type III societies that capture all starlight from a galaxy.  One possible reason is that shrouding a star in a megastructure is prohibitively expensive.  Most of a galaxy's starlight comes from bright stars, which would require structures even larger than the classical Dyson sphere to enclose.  Using a custom spectral synthesis code, I calculate what happens to the spectrum and colors of a galaxy when only stars below a luminosity $L_{\rm min}$ are cloaked.  I find the photometric signatures of galaxies with $L_{\rm min} \le 1\ \Lsun$ are minor, especially for blue, galaxies \vIIChange{with continuing star formation}.  Larger luminosity thresholds ($\ga 30\ \Lsun$) result in galaxies with unnatural colors and luminosities.  \vIIChange{Galaxies} observed in NIR and galaxies \vIIChange{without recent star formation} observed at UV-NIR wavelengths become redder than uncloaked galaxies as $L_{\rm min}$ increases.  \vIIChange{Recently star-forming} galaxies get bluer in UV and blue light when they are cloaked, with colors similar to quasars but very low luminosities.  By selecting on color, we may find Type III societies in large photometric surveys.  I discuss how different metallicities, ages, and initial mass functions affect the results.
\end{abstract}


\section{Introduction}
The search for astroengineering is growing into a prominent branch of the broader Search for Extraterrestrial Intelligence (SETI; \citealt{Tarter01}).  Astroengineering involves the deliberate manipulation of matter and energy on astronomical scales.  These scales can be planetary, stellar, galactic, or even intergalactic, corresponding to Types I, II, III, and IV on \citet{Kardashev64}'s famous scale.  The classic picture of astroengineering is the Dyson sphere, proposed to be either a solid shell or swarm of space stations surrounding a host sun and consuming all of its energy \citep{Dyson60,Badescu06}.  

As our catalogs of galaxies grow and our understanding of their evolution deepens, there is more attention on the possibilities of Type III societies.  These scenarios necessarily require interstellar travel of some kind, in addition to the ability to re-engineer each solar system.  Questions about whether these technologies are feasible, and whether any astroengineering program can be sustained for a long time, spark fierce debate in the literature \citep{Hart75,Tipler80,Sagan83,Brin83,Cirkovic09-Fermi,HaqqMisra09,Wright14-SF}.  But Type III societies use so much energy and alter galaxies so deeply that they could be detected at cosmological distances.  Thus, our effective reach is quadrillions of stars.  Just as traditional SETI can survey many more star systems (e.g., $10^6$ in the Breakthrough Listen survey\vIIIChange{; \citealt{Worden17}}) than searches for extraterrestrial life, while hinging on intelligence commonly evolving from life, so astroengineering searches extend traditional SETI in exchange for additional uncertainty.

Type III societies might develop for a variety of reasons: perhaps resource consumption, ensuring a society's long-term survival, intergalactic communication, enormous science experiments, or stabilization of a galactic environment \citep{Kardashev64,Kardashev85,Cirkovic06-Dysonian,Wright14-SF,Lacki15,Lacki16-K3}.  Four scenarios for Type III astroengineering have been advanced in the literature.  First, they can use large amounts of power to produce non-thermal radiation that we can detect.  The typical example is the radio beacon designed to be found by astronomers within millions of light years \citep{Horowitz93}, but this can also take the form of high energy radiation from pulsars or X-ray binaries, modulated to act as a beacon, or even as pollution from particle accelerators \citep{Chennamangalam15,Lacki15}.  Second, they can move stars and gas around within galaxies, making the galaxy look strange to us \citep{Badescu06,Carrigan12,Voros14}.  Third, they may use their abilities to launch intergalactic travelers, spreading seeds across the entire Universe -- in effect, the Type III society is just a base to build a final Type IV society \citep{Kardashev85,Armstrong13,Olson15,Hooper18}.

Fourth, and most thoroughly studied, the primary manifestation of a Type III society could be its use of a galaxy's luminosity, particularly its starlight.  One way to do this is to put a Dyson sphere around every star in a galaxy \citep[and later papers]{Annis99}.  Another is to capture the radiation with a pervasive screen in the form of interstellar dust \citep{Lacki16-K3}.  Galaxies shrouded in these ways will appear optically faint because of the missing starlight.  Yet any dissipative use of the power (like running a heat engine or doing irreversible computation) will produce waste heat, likely in the form of an infrared or microwave glow, assuming they are bound by known physics \citep{Kardashev85,Wright14-SF,Garrett15,Lacki16-K3}.  Already, infrared waste heat has been the main tracer sought for individual Type II societies within the Galaxy \citep{Sagan66,Slysh85,Criswell85,Timofeev00,Jugaku04,Carrigan09}.  

Indeed, there have been several surveys that search for engineered galaxies that are either optically faint or infrared bright.  A galaxy can be verified to be optically faint with the Tully-Fisher relation, which relates the brightness of a spiral galaxy with the motions of its stars \citep{Tully77}.  These surveys, which have had negative results, have constrained Type III societies to less than 1 in 1,000 galaxies \citep{Annis99,Zackrisson15}.  

\vIIIChange{Waste heat from Type III societies has been sought} in mid-infrared (MIR; characteristic of $300\ \Kelv$ habitable Dyson spheres) \vIIIChange{and} microwaves (characteristic of very cold $3\ \Kelv$ smart dust), \vIIIChange{based on the proposed temperatures of megastructures \citep{Kardashev85,Wright14-SF,Lacki16-K3}.}  They also will appear to lie off the far-infrared radio correlation \citep{Garrett15}, which holds for star-forming galaxies and has only a factor $\sim 2$ scatter \citep{Condon92,Yun01}.  Potentially, these could find Type III societies among millions of galaxies.  One of the most thorough surveys of this form has been Glimpsing Heat from Alien Technologies (GHAT or \GHAT), which looked for extended MIR emission in galaxies observed by the Wide-field Infrared Survey Explorer \citep[WISE][]{Wright14-Results,Griffith15}.  \GHAT~found no signs of a Type III society capturing $\ge 85\%$ of the host galaxy's starlight in an estimated $100,000$ galaxies.  They set weaker limits on Type III societies that capture only some of a galaxy's starlight \citep{Griffith15}.

These results are impressively constraining, and seem to have extreme implications for technological advancement and/or the prevalence of aliens.  But it is important to check whether there are any loopholes, and there are at least two.  First, would waste-heat appear in a visible form?  Maybe the aliens maintain their artifacts at a non-Earthly temperature -- if it's between $\sim 10 \endash 100\ \Kelv$ \vIIIChange{(glowing in far-infrared)} or $\ga 600\ \Kelv$, the waste heat would not have been found yet (\citealt{Lacki16-K3}; see also \citealt{Bradbury00,Osmanov18}).  Perhaps they broadcast it in neutrinos or some other nigh-undetectable particle, or beam it anisotropically away from us\footnote{Beaming is constrained by conservation of etendue, and ultimately thermodynamics.  Basically, the beaming structure (a mirror, a lens) must be proportionally larger than the emitting structure as the solid angle of emission decreases.  This is possible when beaming starlight, by using a Dyson sphere as a mirror -- indeed, it is the idea behind a Shkadov thruster \citep{Badescu06} -- but becomes more difficult when trying to beam waste heat of entire Dyson spheres, or the galaxy as a whole.}, or dump it into black holes.  Or maybe they are somehow storing the energy without dissipating it.

A second objection deals with the extrapolation of a Dyson sphere around one star to Dyson spheres around every star in a galaxy.  A ``classic'' Dyson sphere, consisting of habitable structures, would use the entire mass of Jupiter to build \citep{Dyson60}.  Many have expressed incredulity that something that big could be built, or economical \citep[beginning with][immediately following Dyson's paper]{Maddox60}.  But not all stars have the same luminosity: at a given habitable temperature, the Dyson sphere area increases proportionally with luminosity.  So, where would the builders get the materials to build a Dyson sphere around a red giant or a blue dwarf, with a luminosity $\ga 1,000\ \Lsun?$.  A realistic construction material would mostly contain elements other than hydrogen and helium, limiting the potential of a star to act as a mine.  Even if we suppose the builders can mine more matter from massive stars \citep[as in][]{Criswell85}, blue dwarfs have much smaller mass-to-light ratios than the Sun, and red giants are even worse, with high luminosities but masses comparable to the Sun.  Yet, the vast majority of starlight from galaxies comes from these brilliant stars.

Of course, the screens don't have to take the form of classic Dyson spheres.  Much less massive structures can be built from photovoltaic panels \citep{Bradbury00} or microscopic antennas -- although that does suggest their temperature doesn't have to be Earthlike \citep{Lacki16-K3}.  But another \vIIIChange{search strategy} is to take these as true limits: what happens if only the fainter stars in a galaxy are cloaked in Dyson spheres?  How much would a galaxy dim if only stars fainter than $1\ \Lsun$ were shrouded -- or $0.01\ \Lsun$ or $100\ \Lsun$?  Would its color change significantly?  One advantage of looking for these changes in direct starlight is that it doesn't matter what form (or whether) the waste heat comes out, answering the first objection as well.

Simply doing a deep census of the stellar population within other galaxies could directly find these partially cloaked galaxies, even if the shrouded stars are faint dwarfs.  This strategy is impractical much beyond the Milky Way's satellite system, though, because the individual stars are too faint to be observed and suffer confusion.  For example, the most thorough coverage of M31 is in the Panchromatic Hubble Andromeda Treasury, which achieved $m \approx 28$ ($M \approx 0$) depth in its outer disk \citep{Dalcanton12}.  Even if every M, K, and G dwarf ($M_V \ga 5$; \citealt{Pickles98}) in M31 is shrouded, there would be no sign of it in PHAT.  PHAT has even lower depth in M31's inner disk and bulge, because the stellar fields are crowded enough that stars are blended together \citep{Dalcanton12}.

Outside of the Local Group, stellar censuses are shallower still; in the M81 group, HST images with a depth of $m \approx 28$ would only detect stars with $M \la 0$, like Vega \citep{Dalcanton09}.  Compare this with the $137$ galaxies investigated in \citet{Annis99}, the earliest systematic search for galaxies with missing starlight.  Searches for partial Type III societies are thus still reliant on measurements of integrated light.  \vIIChange{Spectroscopic techniques for determining the abundance of dwarf stars in a galaxy could be useful \citep{Conroy12}, but remain somewhat controversial in that they imply bottom-heavy IMFs in massive elliptical galaxies \citep{vanDokkum10}, in dispute with X-ray binary and lensing models \citep{Smith13,Peacock17}.}  I therefore focus on \vIIChange{integrated} photometric signatures in this paper.  

The goal of this paper is to calculate the effects of cloaking only a part of a galaxy's stellar population on its brightness and colors, using luminosity or luminosity-to-mass ratio as a threshold.  In Section~\ref{sec:Method}, I describe the spectral synthesis calculations I performed, modeling stellar populations that are missing their faint (or bright) stars.  An overview of the effects of partial cloaking on the integrated spectrum of a stellar population is given in Section~\ref{sec:Spectra}.  Then, Section~\ref{sec:Photometry} presents the model results for the photometry of a partially cloaked galaxy.  These include the tracks galaxies trace on color-magnitude and color-color diagrams as stars of greater luminosities are shrouded.  Section~\ref{sec:Conclusion} provides a summary of the results and further possibilities for how a stellar population may be engineered.

\section{Spectral synthesis method}
\label{sec:Method}
A spectral synthesis code able to simulate the screening of stars below a luminosity threshold is necessary to calculate a partially cloaked galaxy's spectrum.  Since more advanced extant codes do not have this feature, I wrote a custom code.  \vIIIChange{This code will be available from the UC Berkeley SETI program's Github site.\footnote{https://github.com/UCBerkeleySETI}  The simulated spectroscopic and photometric data will also be made available.}

First, the code calculates a population distribution for the stars in the simulated galaxy, parameterized by age and initial stellar mass.  This distribution is a combination of the initial mass function (IMF), describing which fraction of stars are born with a mass $M_{\star}$, and a star-formation history (SFH), describing how many stars were born a time $t$ ago.  I assume the IMF doesn't change with time, so the population distribution is:
\begin{equation}
\frac{d^2 N}{dM_{\star} dt} = \frac{dP}{dM_{\star}} \frac{dN_{\star}}{dt} = {\rm IMF} (M_{\star}) \times \frac{{\rm SFH} (t)}{\mean{M_{\star}}}.
\end{equation}
I employed a \citet{Chabrier03} IMF for my baseline calculations, although I also consider a bottom-heavy IMF as well, motivated by claims of the latter IMF in elliptical galaxies \citep[as in][]{vanDokkum10,Spiniello14,MartinNavarro15,LaBarbera17}.  The bottom-heavy IMF used in my calculations:
\begin{equation}
\frac{dP}{dM_{\star}} = \left[\frac{dP_{\rm Chabrier}}{dM_{\star}} (1\ \Msun)\right] \left(\frac{M_{\star}}{1\ \Msun}\right)^{-3.0},
\end{equation}
has a surplus of low mass, low luminosity dwarf stars, and so the partial cloaking of a galaxy has an increased effect.  I normalize it so that the number of $1\ \Msun$ stars is equal to its value for the Chabrier IMF, because the red giants around this mass dominate the light of modern \vIIChange{quiescent} galaxies.  

\begin{figure}
\centerline{\includegraphics[width=9cm]{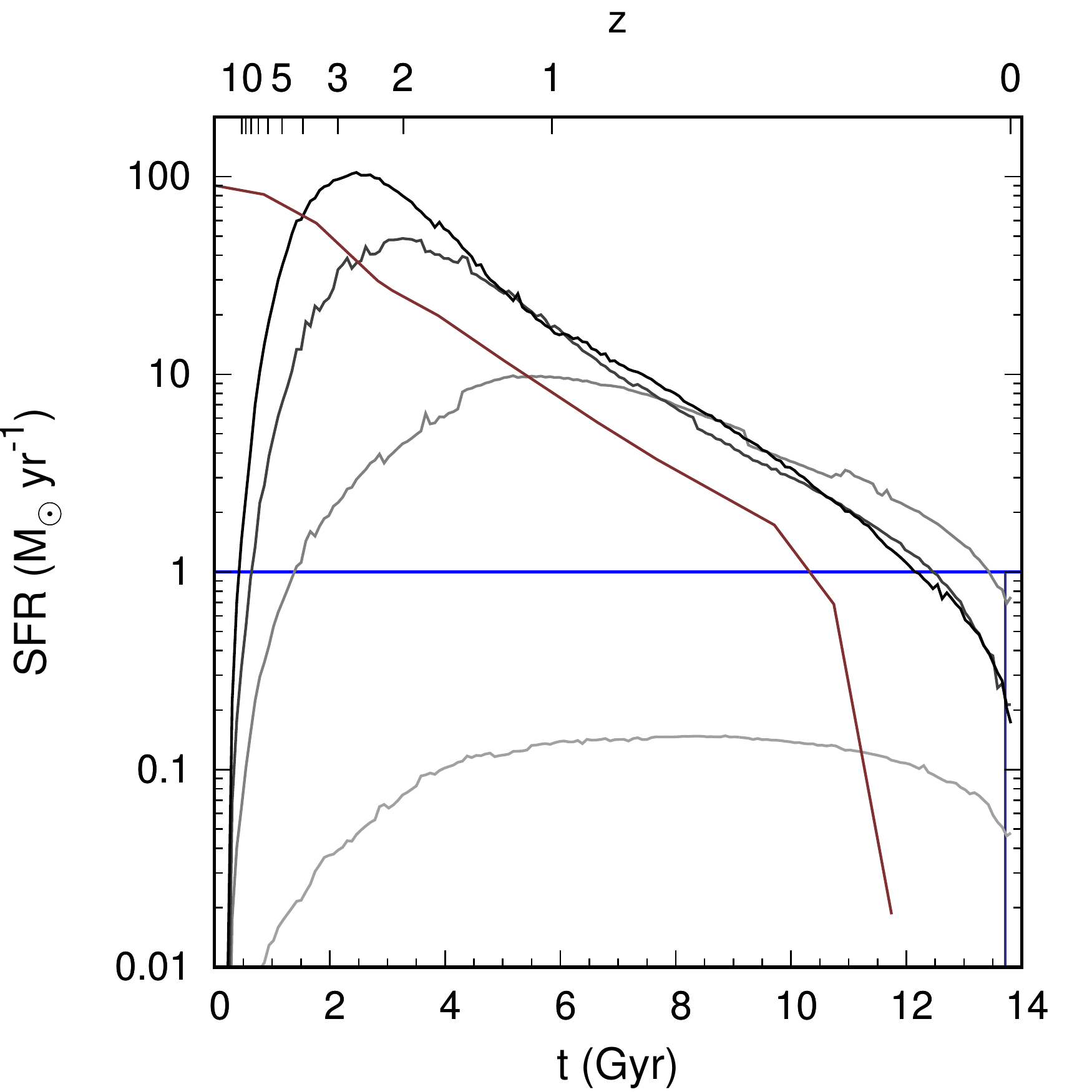}}
\figcaption{\vIIIChange{Star-formation rates} used in this work: a constant SFH \vIIIChange{with age 100 Myr (dark blue) and 13.5 Gyr (bright blue)}; the B13 SFHs for halo masses of $10^{11}$, $10^{12}$, $10^{13}$, $10^{14}\ \Msun$ (light grey to dark grey, respectively); and the M15 SFH \vIIIChange{for $M_{\rm JAM} = 10^{11} \endash 10^{11.5}\ \Msun$} (dark red).\label{fig:SFH}}
\end{figure}

\vIIChange{\vIIIChange{A galaxy's stellar} luminosity distribution depends strongly on the amount of recent star formation compared to stellar mass (proportional to specific star formation).  I consider eight SFHs (Figure~\ref{fig:SFH}) that bracket the range of qualitative behaviors that are encountered in the bulk of main-sequence and quiescent galaxies.  These span a range of luminosity-weighted stellar ages from $1$ to $13\ \Gyr$, with specific star formation rates decreasing from $0.2$ to $0\ \Gyr^{-1}$.  At one extreme are galaxies whose star formation has not slowed down, called \emph{continuing-SF} (star formation) galaxies here.  They are typically bluer galaxies with late-type morphology.  They are represented by a constant SFH.}  \vIIIChange{I focus on two ages for this population: 100 Myr, representative of an ongoing starburst, and 13.5 Gyr, for a normal star-forming galaxy.}\footnote{I \vIIIChange{also} include\vIIIChange{d} a time-averaged version of the mean empirical SFH that \citet{Weisz14} derived for Local Group dwarf irregular (dIrr) galaxies, in which the SFR increases \vIIChange{during the past several Gyr}.  \vIIIChange{This model population behaved similarly with respect to cloaking as the old constant SFR galaxy.}}

Most galaxies have ongoing star formation at a level much smaller than at $z \approx 1 \endash 2$: these are dubbed here \emph{declining-SF} galaxies.  I represent declining-SF galaxies with the simulated SFHs from the simulations in \citep[B13]{Behroozi13} for galaxies in dark matter halos with total masses $10^{11}$, $10^{12}$, $10^{13}$, and $10^{14}~\Msun$.\footnote{\vIIChange{I use these models, which are quantified by total rather than stellar mass, because they are easily available.  They correspond to stellar masses of $10^{8.8}$, $10^{10.4}$, $10^{10.9}$, and $10^{11.1}\ \Msun$, respectively.}}  

\vIIIChange{Lastly}, most stellar mass at $z = 0$ is now located in galaxies with no star-formation, referred to as \emph{quenched} galaxies in this paper.  I consider two quenched galaxy SFHs: (1) an instantaneous burst of age \vIIIChange{$13.5\ \Gyr$ and initial mass $10^{11}\ \Msun$;} and (2) the SFH that the ATLAS3D project derived for $M_{\rm JAM} = 10^{11} \endash 10^{11.5}\ \Msun$ early-type galaxies\vIIChange{, for which there is a spread in stellar ages.} \citep[M15]{McDermid15}.\footnote{\vIIChange{JAM stands for Jeans Anisotropic Multi-Gaussian Expansion, a dynamical model of galaxies, and $M_{\rm JAM}$ is approximately the stellar mass of a galaxy \citep{Cappellari13}.  \vIIIChange{M15} derived additional SFHs for other JAM masses, but their behavior would fall in the range spanned by the declining-SF and the pure burst SFHs.}}  \vIIIChange{The SFHs in M15 are not physical star-formation rates, but the rate the present-day stellar mass built up.  I converted to SFR using the prescription of \citet{Leitner11} for a Chabrier IMF.} \vIIChange{Of course, the general behavior of the SFHs can apply to smaller populations like star clusters.  Throughout the paper, I use the constant SFH as representative of a continuing-SF galaxy and the M15 SFH as a fiducial quenched galaxy, \vIIIChange{in both cases with age 13.5 Gyr}.}

Next, isochrones describe the basic properties (like luminosity, size, and surface temperature) of stars with a specified age as a function of initial mass.  I employ the \software{CMD 3.0} isochrones that are available online \citep{Bressan12,Marigo17}.\footnote{At http://stev.oapd.inaf.it/cgi-bin/cmd}  My models use a grid of stellar ages with \vIIIChange{$4.0 \le \log_{10} t \le 10.13$}, with the upper limit being the oldest available population from the \software{CMD} web interface.\footnote{\vIIIChange{I used the default Reimers mass loss parameter value of $0.2$.}}  The age increases in $\log_{10} t$ steps of $0.01$\vIIIChange{, fine enough to resolve ``AGB boosting'' at 1.6 Gyr \citep{Girardi13}.}  The stellar population also depends on the metallicity of the stars, the abundance of elements heavier than hydrogen and helium.  Old stars and low-mass galaxies have low metallicity, while young stars and high-mass galaxies tend to have high metallicity \citep[e.g.,][]{Timmes95,Tremonti04}.  The metallicities I chose were $0.1\ \Zsun$, $1\ \Zsun$, and $2\ \Zsun$.  \vIIIChange{These isochrones include stars in pre main sequence to the Therally Pulsating Asymptotic Giant Branch (TP-AGB) or carbon burning phases, but they do not include post-AGB evolution \citep{Bressan12,Chen15,Marigo17}.}\footnote{\vIIIChange{I ignored the single post-AGB data point terminating some of the isochrones, because it's only there for compatibility with some codes and its luminosity is deliberately set to $\sim 0$, as stated in the CMD 3.0 FAQ.}}

\begin{figure*}
\centerline{\includegraphics[width=9cm]{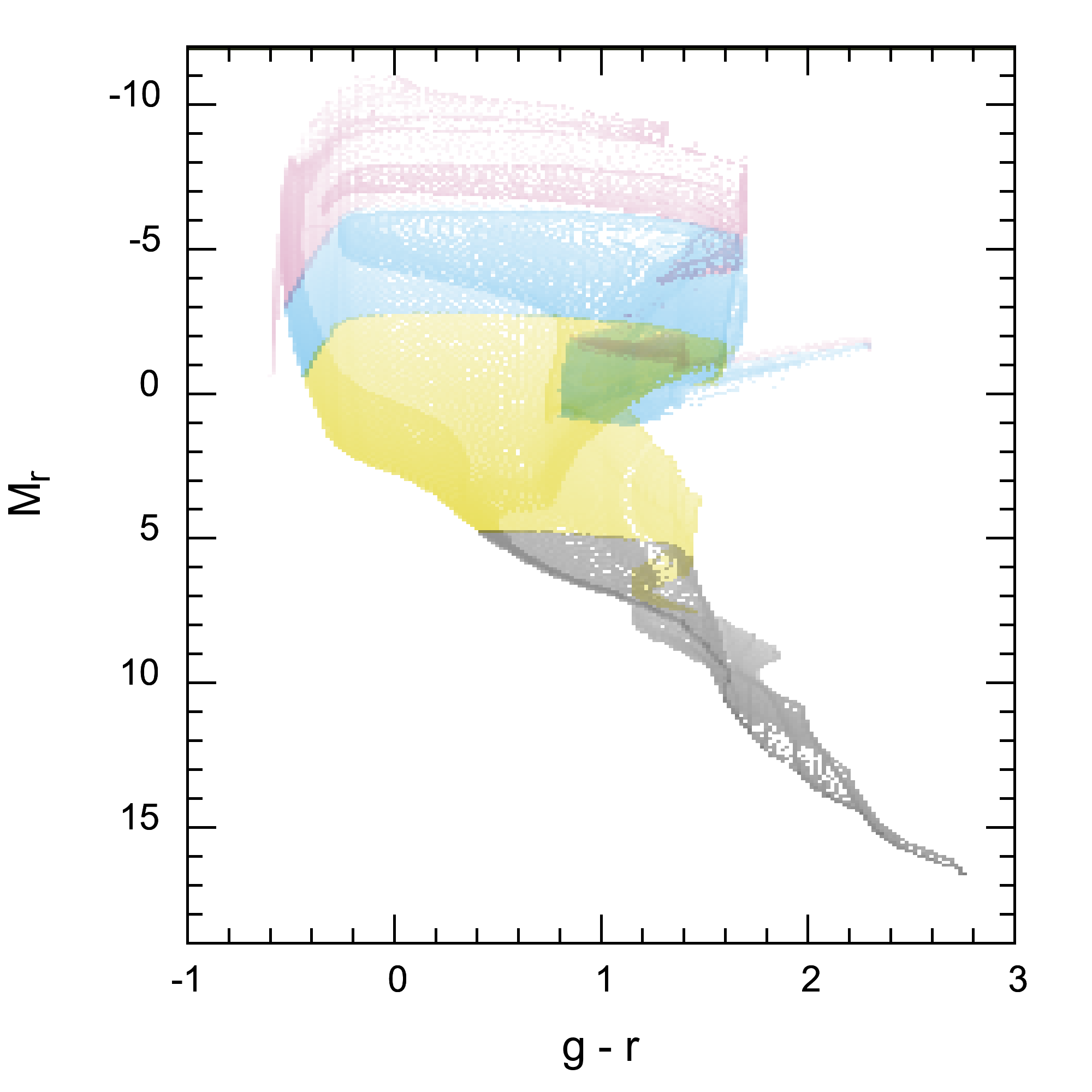}\includegraphics[width=9cm]{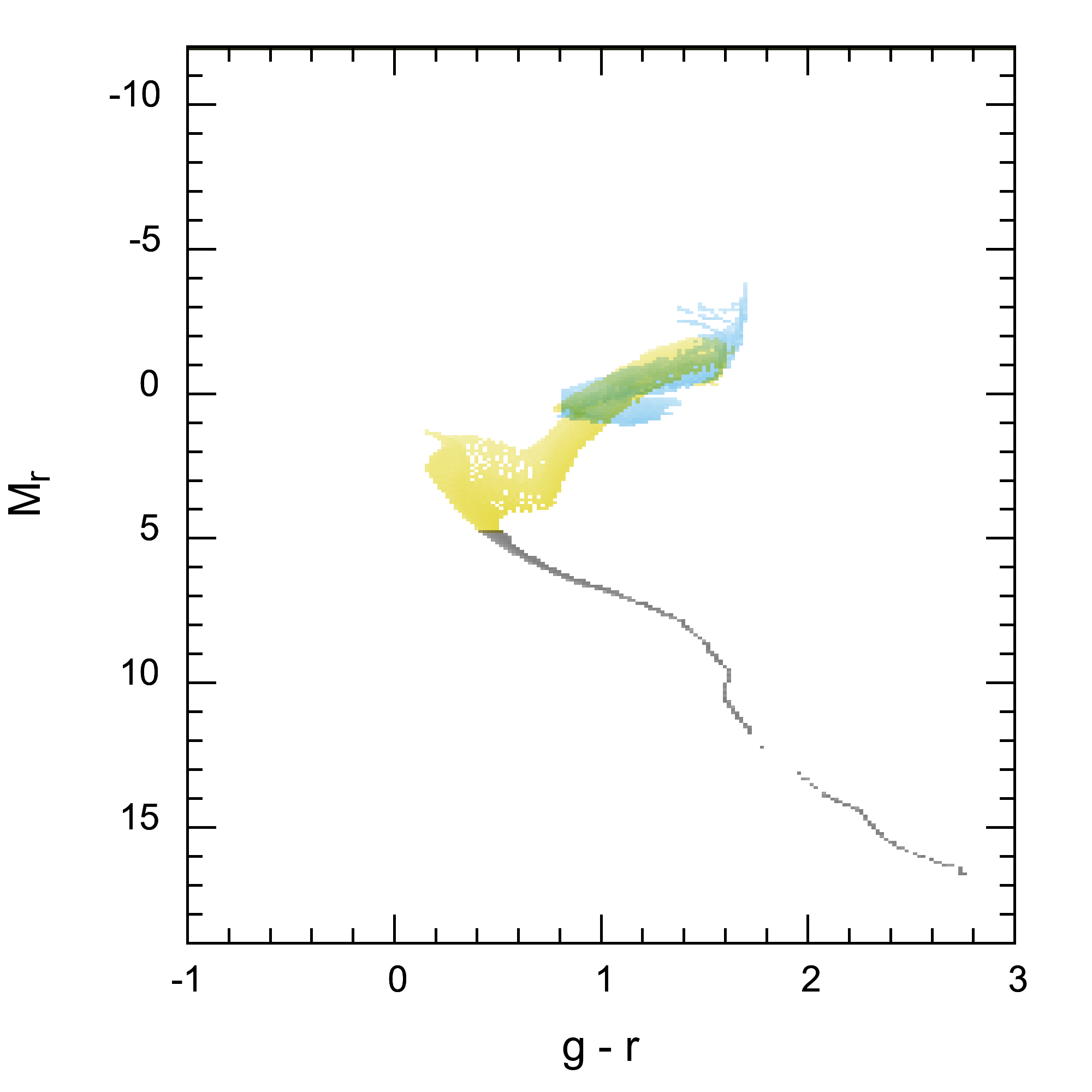}}
\figcaption{Color-magnitude diagrams generated for stellar populations with my code, and the effects of only including stars \vIIChange{within chosen luminosity ranges: $L \le 1\ \Lsun$ (black), $1 \le L \le 10^3\ \Lsun$ (\vIIIChange{yellow}), $10^3\ \le L \le 10^{4.5}\ \Lsun$ (\vIIIChange{sky blue}), and $10^{4.5} \le L$ (\vIIIChange{reddish-purple}).}  The shown diagrams are for a \vIIChange{representative continuing-SF (left) and quenched (right) SFH.}.  Darker shading indicates more stars, using a logarithmic scale.\label{fig:HR}}
\end{figure*}

Finally, the spectrum of the galaxies require the spectra of individual stars in its population.  The BaSeL models of \citet{Lejeune97} provide the spectrum of radiation flux (specific luminosity per unit area) for stellar atmospheres with a wide range of temperature, metallicity, and surface gravity.  Surface gravity parameterizes the pressure in the stellar atmosphere; it is large for dwarf stars like the Sun and small for giant stars.  These flux spectra (denoted $F_{\nu}$) are converted into luminosity spectra $L_{\nu}$ by assuming stars are spherical with $L_{\nu}^{\star} = 4 \pi R_{\star}^2 F_{\nu}$, using the isochrone-supplied radii $R_{\star}$. In some stars from the isochrones, the surface gravity of a star was greater (lesser) than available in the grid of BaSeL spectra for a temperature bin, so I used the flux spectrum for the greatest (smallest) available surface gravity.  There were also a few cases of stars hotter than any of the model spectra\vIIIChange{.  The code models these with blackbody spectra.}

From these ingredients, I calculate the total luminosity spectrum of a model galaxy as
\begin{equation}
L_{\nu} ({\rm galaxy}) = \int \int \frac{d^2 N}{dM_{\star} dt} \times L_{\nu}^{\star} (t, M_{\star}, Z_{\star}) \times \Theta (L_{\star}, M_{\star} dM_{\star}) dt.
\end{equation}
$\Theta$ is the threshold factor, which is $0$ for stars that are screened after passing a specified luminosity (or luminosity-to-mass) cut, and $1$ otherwise. \vIIIChange{The code also allows for cuts on mass and by (approximate) stage of stellar evolution.}

\begin{deluxetable*}{llccc}
\tablecolumns{5}
\tablecaption{Photometry Systems Used\label{table:PhotometrySystems}}
\tablehead{\colhead{System} & \colhead{Filters} & \colhead{Type} & \colhead{Normalization} & \colhead{Reference}}
\startdata
FUSE                & FUV, NUV                        & AB   & Given in reference          & \citet{Morrissey05}\\
Johnson-Cousins     & $UBVRI$                         & Vega & {\tt alpha\_lyr\_stis\_008}  & \citet{Bessell90}\\
SDSS                & $ugriz$                         & AB   & $3631\ \Jy$                 & \citet{Doi10}\\
Dark Energy Survey  & $grizY$                         & AB   & $3631\ \Jy$                 & \citet{Burke18}\\
PanSTARRS-1         & $grizyw$                        & AB   & $3631\ \Jy$                 & \citet{Tonry12}\\
LSST                & $ugrizy$                        & AB   & $3631\ \Jy$                 & \citet{LSST09}\\
\emph{Gaia}         & $G$, $G_{\rm BP}$, $G_{\rm RP}$ & Vega & {\tt alpha\_lyr\_mod\_002}  & \citet{Carrasco16,Evans18}\\
2MASS               & $JHK_s$                         & Vega & Given in reference          & \citet{Cohen03,Skrutskie06}\\
UKIRT               & $ZYJHK$                         & Vega & {\tt alpha\_lyr\_stis\_008} & \citet{Hewett06}\\
\emph{Spitzer} IRAC & IRAC-1, -2, -3, -4              & AB   & $3631\ \Jy$                 & \citet{Fazio04}\\
WISE                & W1, W2, W3, W4                  & Vega & Given in reference          & \citet{Wright10}
\enddata  
\tablecomments{The Johnson-Cousins system assigns magnitude 0.03 to Vega in all bands \citep{Bessell05}, and the \emph{Gaia} photometry sets Vega's apparent magnitude at 0.023 \citep{Carrasco16}.\\
I downloaded the LSST filter throughput curves from the LSST Github site at https://github.com/lsst/throughputs/tree/master/baseline.}
\end{deluxetable*}

I also calculate the absolute magnitude of the model galaxy through several filters using this luminosity spectrum.  These magnitude were found using the photon number flux, rather than the energy flux.  Given a transmission function $T_x(\nu)$ for a filter band $x$ (and the atmosphere, if relevant), an absolute magnitude ${\cal M}_x$ in that band is calculated using
\begin{equation}
{\cal M}_x = \displaystyle -2.5 \log_{10} \frac{\displaystyle \int \frac{1}{h\nu} \frac{L_{\nu}}{4 \pi D_0^2} T_x(\nu) d\nu}{\displaystyle \int \dot{N}_{\nu}^0 T_x(\nu) d\nu},
\end{equation}
where $D_0 = 10\ \pc$ and $\dot{N}_{\nu}^0$ is the photon number flux of a source with magnitude $0$.  I list the magnitude systems for which I calculated simulated photometry in Table~\ref{table:PhotometrySystems}.  The references give the transmission curves.  Broadly speaking, the systems are either AB magnitudes, in which the magnitude zero point is given by $F_{\nu} = 3631\ \Jy$, or the Vega magnitude system, in which the zero point is set using the spectrum of Vega, possibly with a small magnitude offset.  For Vega magnitude system photometry, I normalize with the model spectra of \citet{Bohlin04}, which are available online.\footnote{ftp://ftp.stsci.edu/cdbs/current\_calspec}  When discussing results in this paper, I focus on photometry in the SDSS and 2MASS filter systems.  Unless otherwise indicated, $ugriz$ refers to the SDSS passbands.  Two stellar color-magnitude diagrams (CMDs) constructed by a modification of the code are shown in Figure~\ref{fig:HR}, for an old continuing-SF and a quenched galaxy.

Because they are purely stellar models with no chemical evolution, my models have several limitations.
\begin{itemize}
\item \vIIIChange{Dust extinction in the Milky Way and the host galaxy is not modeled.  The Milky Way extinction towards the Galactic Poles is typically around $A_V \approx 0.05 \endash 0.1$ \citep{Schlegel98}.  Extinction in other galaxies depends on their morphology and inclination.  Face-on spiral galaxies at $z = 0$ typically have $\mean{A_B} \approx 0.2 \endash 0.5$, while early-type galaxies have $\mean{A_B} \approx 0.05 \endash 0.1$ \citep{Calzetti01}.  The extinction is much greater in edge-on gas-rich galaxies and galaxies at $z \ga 1$.  Reddening is particularly acute in the ultraviolet, with the majority of UV light absorbed in star-forming galaxies \citep{Lisenfeld96,Bell03}.  The attenuation is patchy, however, because of the density fluctuations in the multiphase ISM.  High mass stars are initially veiled in gaseous star forming regions and are preferentially extincted, but their feedback excavates bubbles in these clouds over Myrs.}
\item Nor do I include interstellar dust emission\vIIIChange{, which }dominates the \vIIIChange{MIR} and \vIIIChange{FIR} SEDs of star-forming galaxies\vIIIChange{\citep{Purcell76,Draine01,Li01,Hwang10}.}
\item \vIIIChange{Although my code calculates the megastructures' thermal emission, I will not discuss the results,} since we do not know what temperature \vIIIChange{they would be.  Temperatures greater than $1,000\ \Kelv$ are necessary to} contribute to the galaxy's SED \vIIIChange{at wavelengths less than $2\ \um$ (around K band).}
\item I do not include nebular emission lines like H$\alpha$ or OIII, which are bright in intensely star-forming galaxies \citep{Charlot01}.  In some cases, the emission lines can be strong enough to give galaxies unusual colors \citep{Atek11}, \vIIIChange{as} with the Green Pea galaxies \citep{Cardamone09}.
\item Since I do not include the effects of circumstellar dust or different abundance patterns in the atmospheres of stars, my models do not treat Thermally Pulsating Asymptotic Giant Branch (TP-AGB) stars properly, although they are included in the PARSEC isochrones \citep{Marigo08,Marigo17}.  These stars can have either oxygen- or carbon-rich atmospheres, \vIIIChange{affecting} their SEDs.  Intermediate mass TP-AGB stars, with ages of $\sim 1\ \Gyr$, may actually dominate the NIR emission of post-starburst galaxies \citep{Maraston05,Maraston06}.  They are particularly significant as a source of MIR emission due to their circumstellar dust shells \citep{Conroy13}.  The treatment of TP-AGB stars is widely considered a difficult but important issue to treat in stellar evolution models \citep{Choi16,Marigo17}, and for spectral synthesis based on those models \citep{Maraston06,Conroy13}.
\item Galaxies do not have a constant metallicity, but generally become more metal-rich with time \citep{Timmes95} and near their centers \citep{Henry99}.  Thus, in reality, there should be a spread in the metallicity in the dwarf star and low mass post-main sequence population.
\item I do not include any treatment of binary star evolution.  Objects that are the result of binary star interactions \vIIIChange{are not modeled.  These include extreme horizontal branch (HB) stars, suspected to be the source of the UV excess in quiescent galaxies below 200 nm \citep{Han07}, and} supersoft X-ray sources \vIIIChange{that contribute to extreme UV luminosity} \citep{Kahabka97}. 
\end{itemize}
For these reasons, the models \vIIIChange{should not be considered accurate at} MIR and longer wavelengths, or in far-ultraviolet or shorter wavelengths.  

\section{Integrated Spectra of Partially Cloaked Galaxies}
\label{sec:Spectra}
\begin{figure*}
\centerline{\includegraphics[width=18.5cm]{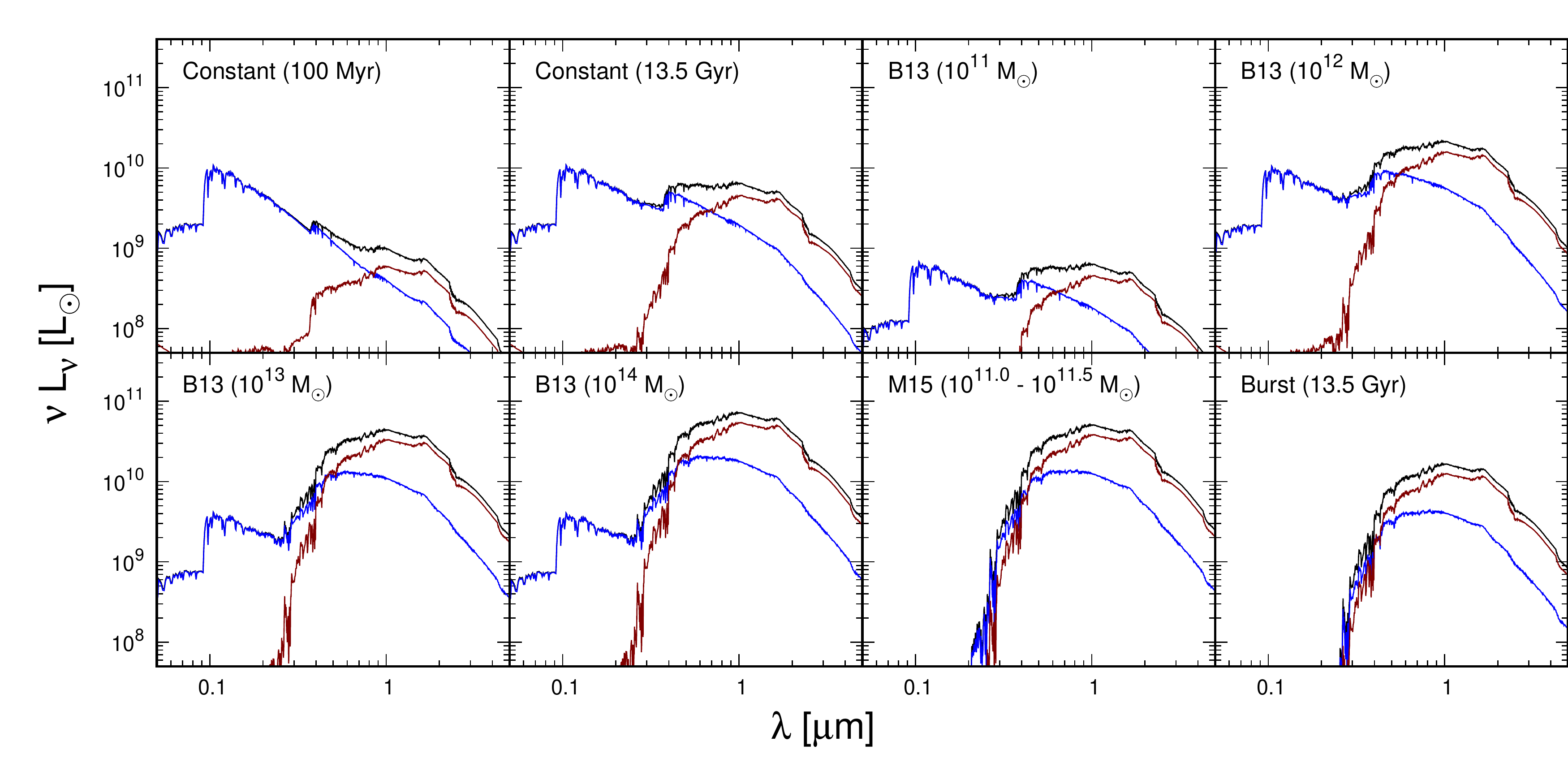}}
\figcaption{\vIIIChange{Stellar spectra of natural galaxies with different SFHs, from \vIIChange{continuing-SF to quenched}.  The contributions to the SED from pre-MS and MS stars (blue) and post-MS stars (red) are also shown.}  In every case, the stellar populations have a Chabrier IMF and a metallicity of $1\ \Zsun$. \label{fig:StageSpectra}}
\end{figure*}

The stellar spectra of natural, uncloaked galaxies qualitatively consist of two peaks (Figure~\ref{fig:StageSpectra}).  One peak reaches its maximum near the Lyman limit, composed of the blue and ultraviolet light from young, massive stars \vIIIChange{on the main sequence (blue line, top)}.  The other peak, mostly consisting of the red and near-infrared light from older red giants \vIIIChange{(red line, top)}, reaches its maximum near $1\ \um$.  The \vIIIChange{blue} peak becomes more prominent as the SFH progresses from \vIIIChange{quenched} to \vIIIChange{continuing-SF} galaxies, \vIIIChange{because the increasing number of high-mass stars contribute more to the luminosity}.  In addition, \vIIIChange{an optical plateau in the SED extending to $\sim 400\ \nm$ is adjoined to the red peak in continuing-SF galaxies.\footnote{\vIIIChange{Note the flat top of the red peak in Figure~\ref{fig:StageSpectra} for these galaxies, compared to the rounded top for the quenched galaxies.}}  Dwarf stars dominate this plateau's luminosity when $\lambda \la 650\ \nm$, and this contribution is mostly from A and F dwarfs with $1\ \Msun \la M \la 3\ \Msun$.}

\begin{figure*}
\centerline{\includegraphics[width=18.5cm]{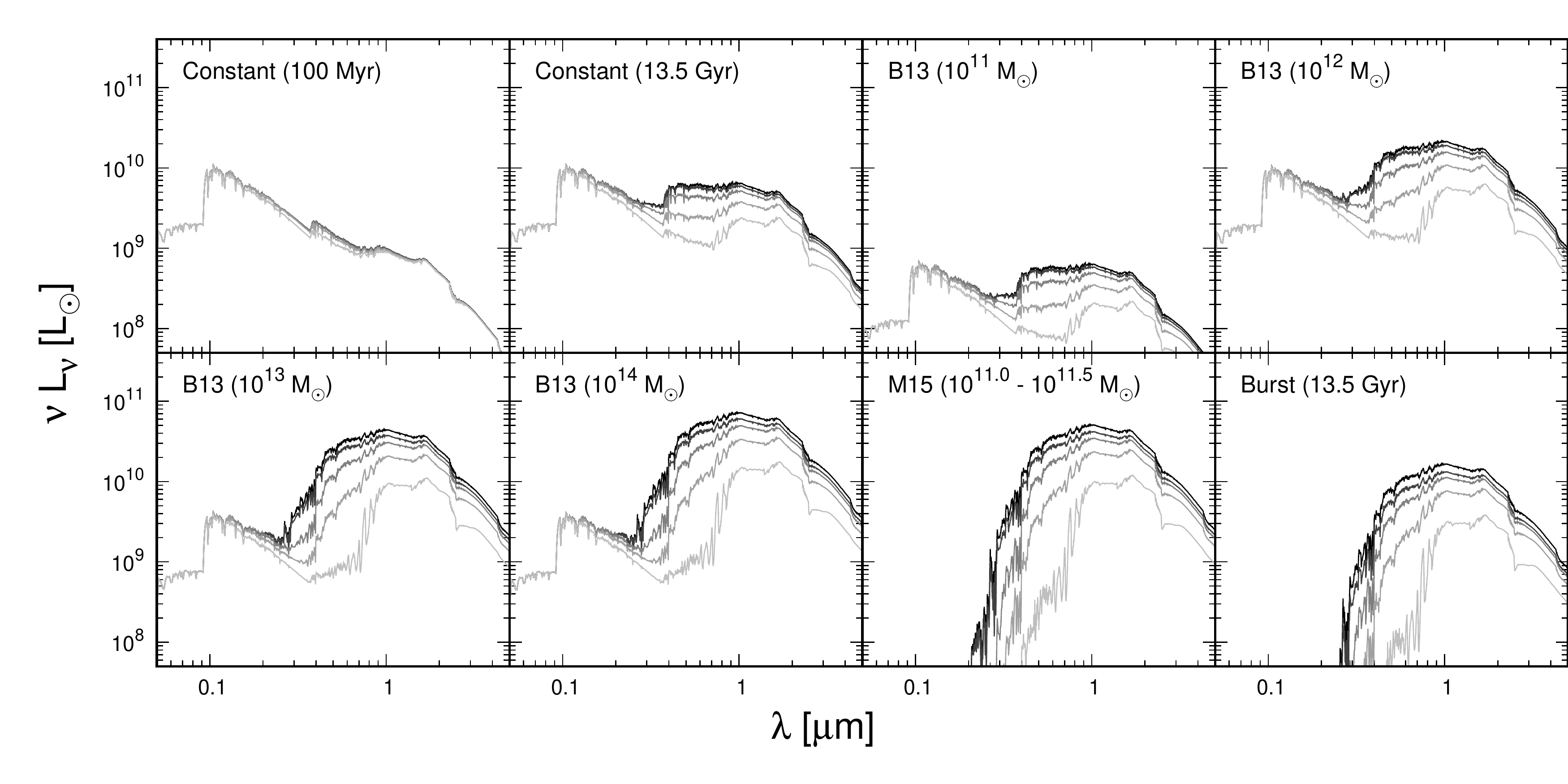}}
\figcaption{Stellar spectra of \vIIIChange{partially} and cloaked \vIIIChange{(bottom)} galaxies for several values of $L_{\rm min}$.  Within each \vIIIChange{subpanel}, from top to bottom and darker to lighter lines, these are \vIIIChange{the unscreened spectra} and then $L_{\rm min} \vIIIChange{=} 1\ \Lsun$, $10\ \Lsun$, $100\ \Lsun$, and $1,000\ \Lsun$. In every case, the stellar populations have a Chabrier IMF and a metallicity of $1\ \Zsun$. \label{fig:Spectra}}
\end{figure*}

\vIIIChange{Screening stars with luminosity below $L_{\rm min}$ (Figure~\ref{fig:Spectra}) changes the shape of the integrated SED.  A small $L_{\rm min} \le 1\ \Lsun$ means only low luminosity (GKM) dwarfs are cloaked.}  \vIIIChange{These stars} are brightest at wavelengths longer than $\sim 400\ \vIIChange{\nm}$, so they contribute mostly to the red peak and \vIIIChange{the optical plateau}.  Screening only these stars has insignificant effects on the ultraviolet luminosity of a galaxy \vIIIChange{(Figure~\ref{fig:Spectra})}.  At visible to near-infrared wavelengths, a screening threshold $L_{\rm min} = 1\ \Lsun$ leads to a \vIIIChange{$< 20\%$} drop in specific luminosity, with a more pronounced drop for \vIIChange{quenched galaxies}.  Thus, galaxies missing these stars would not \vIIChange{have qualitatively different spectral energy distributions} and finding them must require careful analysis.

\begin{figure}
\centerline{\includegraphics[width=9cm]{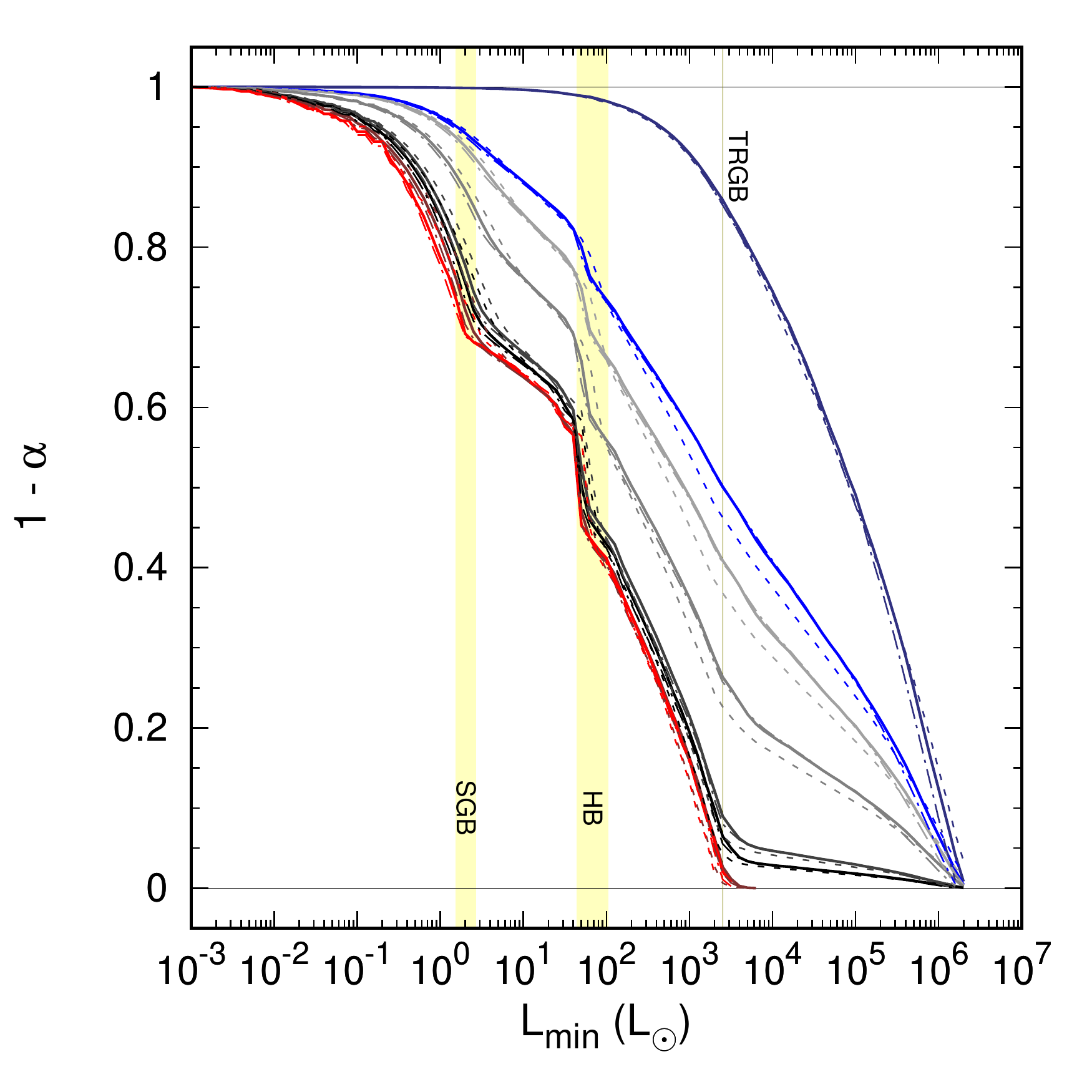}}
\figcaption{How the \vIIChange{unscreened stellar} bolometeric luminosity of a galaxy depends on $L_{\rm min}$.  The different SFHs are \vIIIChange{a 100 Myr old constant SFR galaxy} (top, dark blue), constant \vIIIChange{SFR} \vIIIChange{after 13.5 Gyr} (medium blue), B13 SFHs for halo masses of $10^{11}$, $10^{12}$, $10^{13}$, and $10^{14}\ \Msun$ (from top to bottom, grey), the M15 \vIIChange{quenched} SFH (dark red), and a \vIIIChange{13.5 Gyr old} burst (bottom, medium red).  The solid lines are for a metallicity of $1\ \Zsun$, while the dot--long-dashed lines are for $2\ \Zsun$, and the dotted lines are for $0.1\ \Zsun$.  All shown functions are for a Chabrier IMF and a stellar population age of $13.5\ \Gyr${\vIIIChange, except for the 100 Myr old star-forming population}.  Also marked as yellow bands are the luminosity ranges for the SGB, HB, and TRGB phases for an isochronal population with age \vIIIChange{$13.5\ \Gyr$} and Solar metallicity. \label{fig:MBol}}
\end{figure}

More dramatic changes in a galaxy's starlight occur if $L_{\rm min} \gg 1\ \Lsun$.  Figure~\ref{fig:MBol} depicts the fraction of a galaxy's \vIIChange{stellar} luminosity \vIIChange{that} remains unscreened as a function of $L_{\rm min}$ for various SFHs, denoted $1 - \alpha$ in the \GHAT~AGENT formalism.  When $L_{\rm min} = 1\ \Lsun$, this fraction remains at $80 \endash \vIIIChange{99.9}\%$, but it falls to $64 \endash \vIIIChange{99.7}\%$ for $L_{\rm min} = 10\ \Lsun$, $41 \endash \vIIIChange{98}\%$ for $L_{\rm min} = 100\ \Lsun$, and $16 \endash \vIIIChange{92}\%$ for $L_{\rm min} = 1,000\ \Lsun$.  In all cases, \vIIChange{quenched} galaxies fade more than \vIIChange{galaxies with recent star formation \vIIIChange{because they lack the additional light from high and intermediate-mass stars.}}

The increase in $\alpha$ does not occur steadily as $L_{\rm min}$ increases, but happens in spurts.  These can be related to the luminosity of stars at distinct phases of their evolutions.  The yellow bands in Figure~\ref{fig:MBol} mark the luminosity of stars in the subgiant branch (SGB), horizontal branch or red clump (RC), and tip of the RGB (TRGB) for an isochronal stellar population of age $13.5\ \Gyr$.  In \vIIChange{quenched} galaxies, the first sudden growth in $\alpha$ occurs as $L_{\rm min}$ approaches the main sequence turn-off luminosity, followed by a plateau for the luminosity range occupied during the short-lived SGB phase.  Then, $\alpha$ slowly increases with $L_{\rm min}$ as stars along the red giant branch are cloaked, but there's a sudden jump near $40\ \Lsun$ where nearly all the stars in the red clump (low mass horizontal branch; \citealt{Girardi16}) are shrouded.  Roughly $10\%$ of a \vIIChange{quenched} galaxy's bolometric luminosity is concentrated in the red clump stars.  The increase in $\alpha$ continues until the TRGB is reached \citep[c.f.,][]{Salaris02}, leaving only a few TP-AGB stars providing a residual luminosity.  With \vIIChange{declining- and continuing-SF} galaxies, these features are much more subtle, because much of the luminosity is provided by bluer main sequence stars \vIIIChange{that do not have ``clumps'' or breaks in their luminosity distribution.}
  
\begin{figure*}
\centerline{\includegraphics[width=18.5cm]{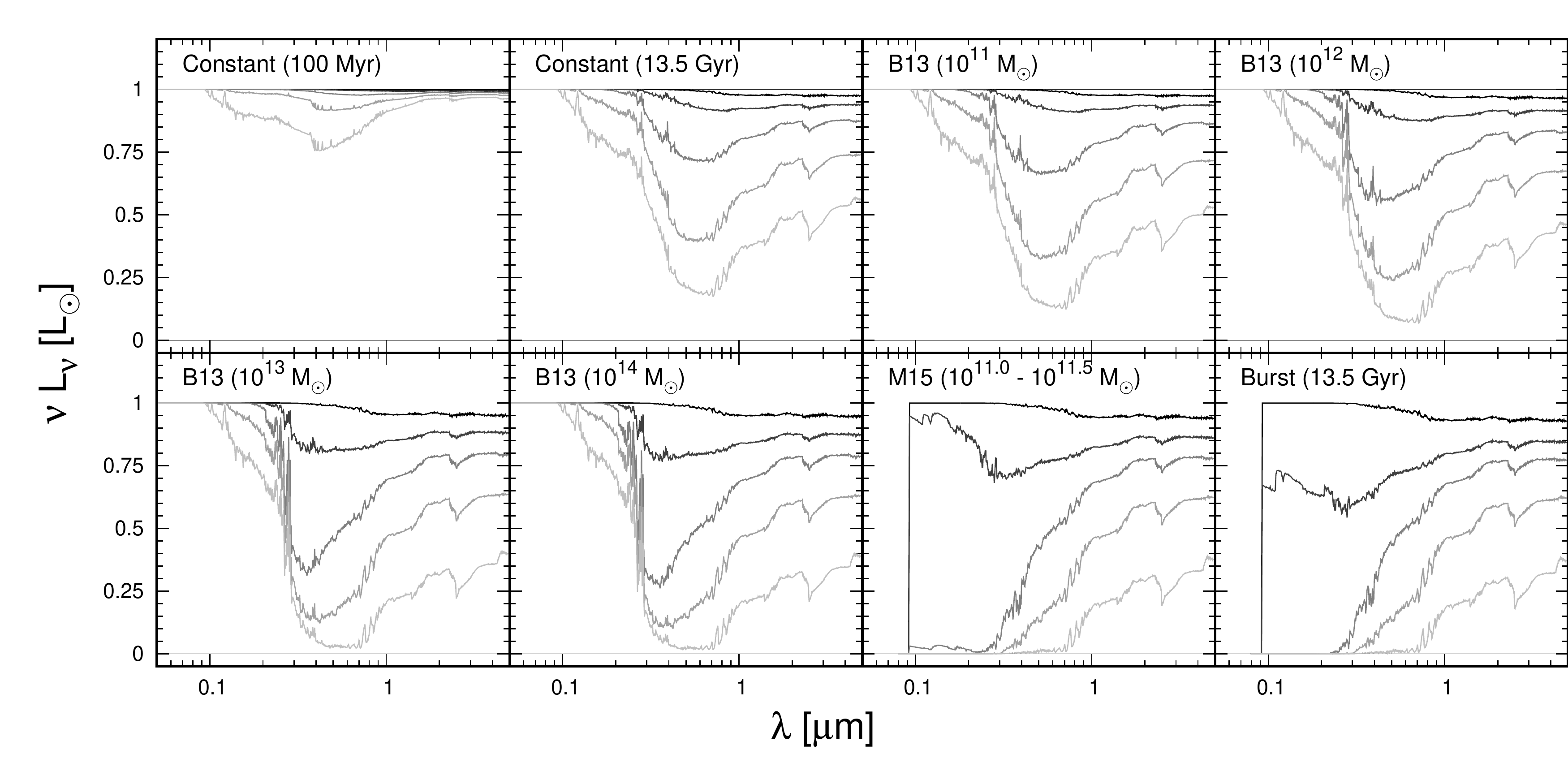}}
\figcaption{The ratio of screened $L_{\nu}$ to the unscreened specific luminosity\vIIChange{, measuring the relative strength of the cloaking effect at each wavelength for each SFH.  A value of $1$ implies none of the galaxy's luminosity at that wavelength is screened, whereas a value of $0$ implies all of it is screened.}  Effects are shown for $L_{\rm min}$ of $0.1 \Lsun$ \vIIIChange{(black)}, \vIIIChange{and} $1$, $10$, $100$, and $1,000\ \Lsun$ (same line styles as Figure~\ref{fig:Spectra}).  \label{fig:RatioSpectra}}
\end{figure*}

As seen in Figure~\ref{fig:Spectra}, increasing $L_{\rm min}$ up to $\sim 1,000\ \Lsun$ erodes the red peak while leaving the blue peak mostly untouched.  The fading is most significant in the visible parts of the spectrum ($300 \endash 800\ \nm$), as demonstrated when the ratio of the screened and the unscreened spectra are plotted (in Figure~\ref{fig:RatioSpectra}).  For the flat SFH galaxies, \vIIIChange{as} $L_{\rm min}$ \vIIIChange{grows from} $10$ \vIIIChange{to} $1,000\ \Lsun$, \vIIIChange{A and F dwarfs are cloaked.  The optical} plateau \vIIIChange{adjoined to} the red peak \vIIIChange{is screened}, \vIIIChange{leaving} the integrated spectrum \vIIIChange{with} two sharp peaks \vIIIChange{from OB dwarfs and bright red giants}.  \vIIIChange{Meanwhile, in the same $L_{\rm min}$ range,} the luminosity at $\sim 300\ \nm$ is virtually extinguished for \vIIIChange{quenched} galaxies.  Yet, even the NIR summit of the red peak is still eroded by screening low-to-mid luminosity stars, if not as quickly because the light from brilliant red giants remain.  In addition, a fairly narrow dip appears in the spectrum around $2.5-2.6\ \um$ as $L_{\rm min}$ grows. It is due to a water molecule absorption band in the spectra of the brightest red giants \citep{Rayner09}.  Unfortunately, it is not covered by standard photometric filters, due to water vapor absorption in Earth's atmosphere, and neither \emph{Spitzer} IRAC nor WISE covered that wavelength region either.  

\begin{figure*}
\centerline{\includegraphics[width=18.5cm]{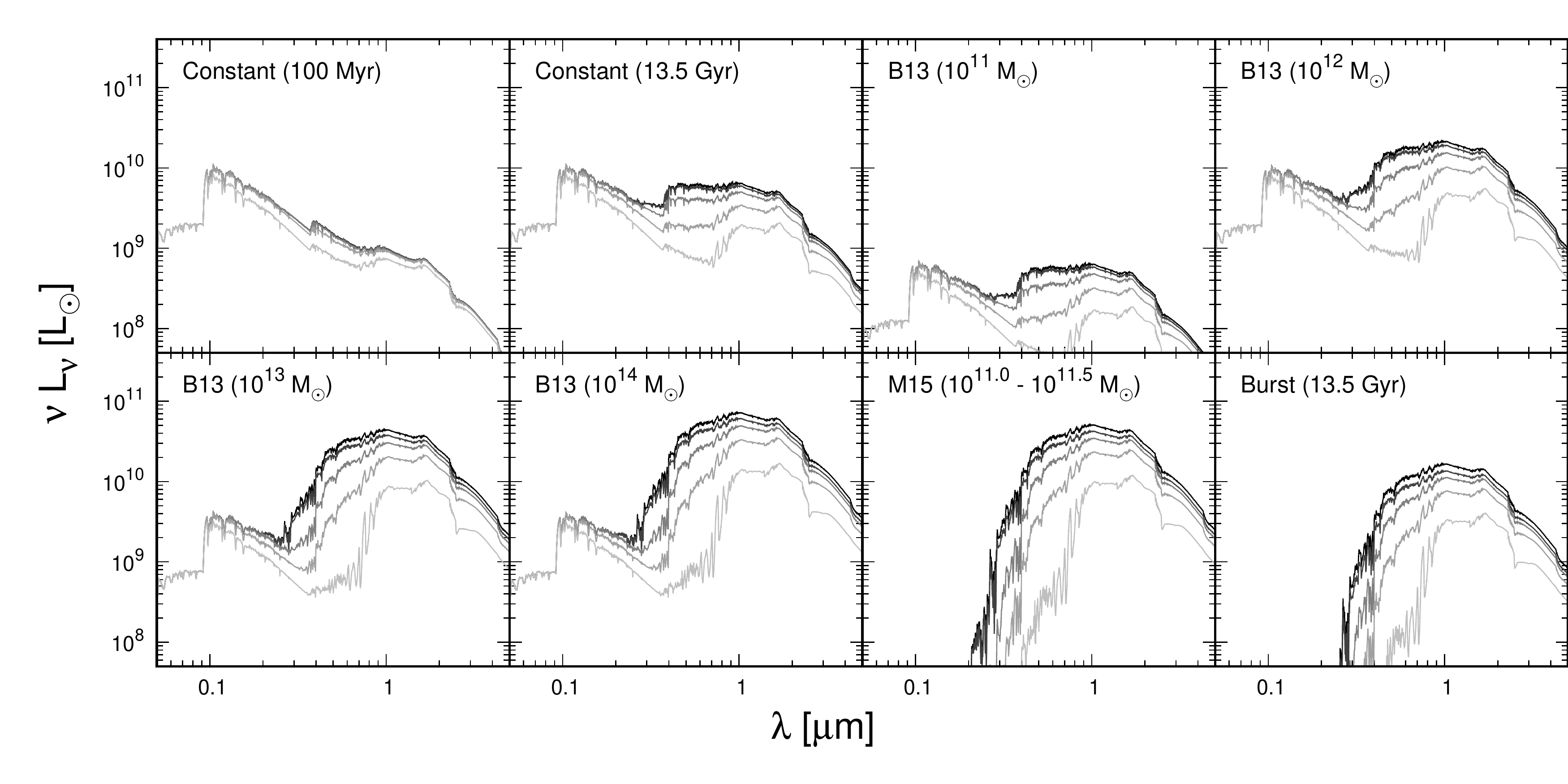}}
\figcaption{Stellar spectra of partially cloaked galaxies with constant SFH where stars below a luminosity-to-mass ratio are cloaked.  From top to bottom, we have $(L/M)_{\rm min} = 0$, $1\ \LMsun$, $10\ \LMsun$, $100\ \LMsun$, and $1,000\ \LMsun$.  \label{fig:LtoMCut}}
\end{figure*}

\begin{figure}
\centerline{\includegraphics[width=8cm]{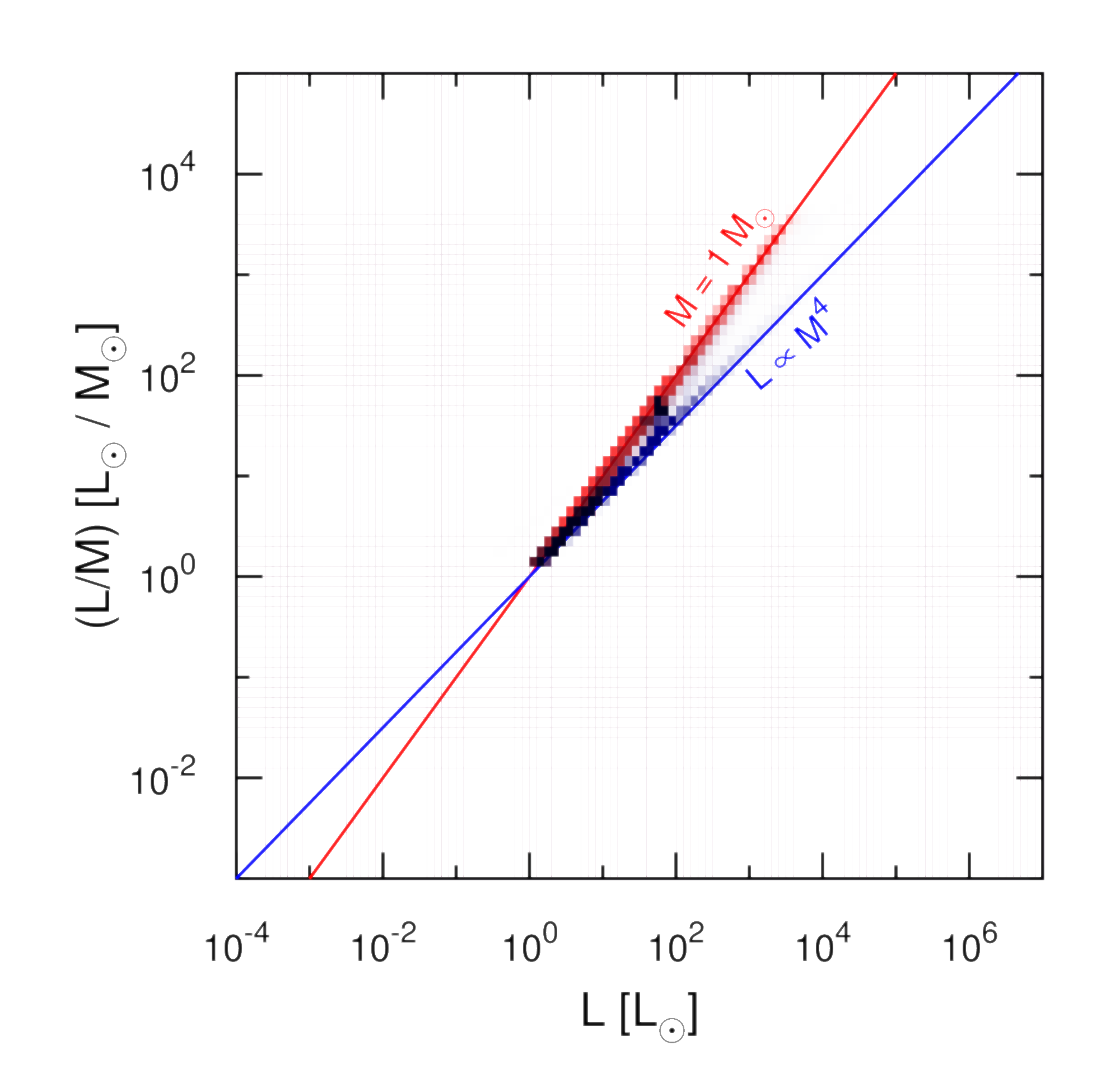}}
\figcaption{The \vIIChange{luminosity-weighted} distribution of stars in $L$ versus $(L/M)$ for \vIIChange{\vIIIChange{an old} continuing-SF galaxy (blue) and a quenched galaxy (red)}.  Main sequence stars mostly fall on $L \propto M^4$ (blue), while \vIIChange{the majority of} post-main sequence stars \vIIChange{have $M \approx 1\ \Msun$}.\label{fig:LtoMTrends}}
\end{figure}

I also considered the effects of using the luminosity-to-mass ratio as a cut rather than the stellar luminosity.\footnote{I used the initial mass of a star when making this cut, which is appropriate if ETIs build megastructures before mass loss sets in.}  More massive stars tend to start with substantial protoplanetary disks and more massive planets \citep[e.g.,][]{Johnson10,Andrews13}, and so have more material around to build a megastructure.  

I found that the changes in the spectra as $(L/M)_{\rm min}$ increases largely looked the same as when $L_{\rm min}$ is used (Figure~\ref{fig:LtoMCut}).  \vIIChange{The reason is simply because there are two distinct kinds of bright stars contributing most of the light, and these groups have different colors.  Furthermore, each population lies on a nearly monotonic relation between $L$ and $L/M$.  As seen in Figure~\ref{fig:LtoMTrends}, the main sequence stars have $L \propto M^4$, while most post main-sequence stars have $M \approx 1\ \Msun$ since low mass stars dominate the IMF.  Thus, whether looking in UV/blue light -- where main sequence stars dominate -- or red/infrared light -- where the red giants dominate -- there's a one-to-one relation between $L$ and $L/M$, and which one is used as the cut does not matter much.  This monotonicity fails when both populations contribute equally to the spectrum (as for continuing-SF galaxies near $0.5\ \um$), resulting in the galaxies being somewhat redder and fainter when $(L/M)_{\rm min}$ is used instead of $L_{\rm min}$.}  \vIIIChange{It can also fail for young stellar populations, including starbursts that are $\ll 100\ \Myr$ old.  Stellar populations of age $\sim 100\ \Myr$ also contain low mass pre-main sequence stars that have not yet settled on the $L \propto M^4$ relation.} 

\section{Photometry of Partially Cloaked Galaxies}
\label{sec:Photometry}
Photometric surveys provide large datasets of galaxy colors and brightnesses \citep[e.g.,][]{AdelmanMcCarthy07,LSST09}.  Since the starlight screened by partially cloaked galaxies is broadband, these surveys may provide an opportunity to quickly search up to several billion galaxies for Type III societies with $L_{\rm min} \gg 1\ \Lsun$.

\begin{figure*}
\centerline{\includegraphics[width=16cm]{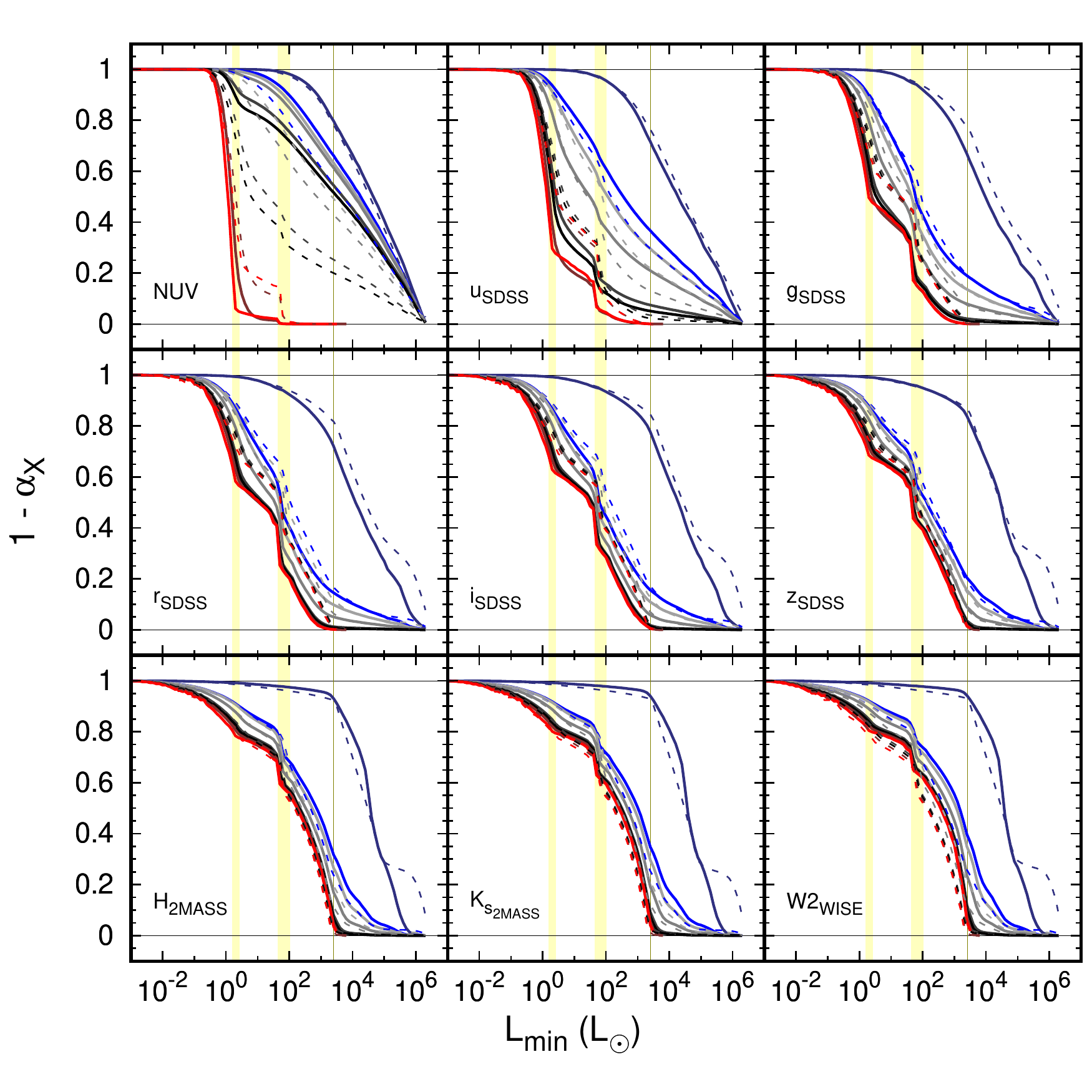}}
\figcaption{The fraction of flux in representative bandpasses that remains unscreened.  Shown are GALEX NUV (top left), $u_{\rm SDSS}$ (top center), $g_{\rm SDSS}$ (top right), $r_{\rm SDSS}$ (middle left), $i_{\rm SDSS}$ (middle center), $z_{\rm SDSS}$ (middle right), $H_{\rm 2MASS}$ (bottom left), $\textrm{K}_{s_{\rm 2MASS}}$ (bottom center), and W2 (bottom right).  Line styles are the same as in Figure~\ref{fig:MBol}.\label{fig:MagnitudeFunctions}}
\end{figure*}

The declining luminosity in various filter bands as $L_{\rm min}$ increases are plotted in Figure~\ref{fig:MagnitudeFunctions}.  I generalize the AGENT parameterization in \citet{Wright14-Results} to introduce the $\alpha_x$ value, the fraction of starlight screened in some band $x$.  

\vIIChange{As described in Section~\ref{sec:Spectra}, the effects of partial cloaking on a galaxy can be understood from the relative contributions of main-sequence and post-MS stars.  The main sequence forms a \vIIIChange{smooth luminosity} distribution, with bright stars being bluer and younger, and it emits a galaxy's light in blue bands (like $u$).  In a continuing-SF galaxy, therefore, increasing $L_{\rm min}$ cause\vIIIChange{s} the dwarf stars' integrated light to smoothly fade and become bluer\vIIIChange{, as $L_{\rm min}$ slides up the main sequence} (Figure~\ref{fig:HR}).  In quenched-SF galaxies, all dwarf stars are cloaked once $L_{\rm min}$ passes the main sequence turn-off, and the flux in blue bands is extinguished precipitously.

Post-MS stars have a more complicated luminosity distribution, resulting in more complicated evolution in redder bands (like $z$) as $L_{\rm min}$ increases.  The red giant and asymptotic giant branches together form a rough sequence with bright stars being redder.  But horizontal branch stars complicate matters, since the low-mass HB stars tend to lie near the red clump, with similar luminosities and colors.  As $L_{\rm min}$ increases, the giant stars' integrated light fades and becomes redder, but there tends to be a sudden drop in flux near $30\ \Lsun$ when red clump stars are cloaked.  \vIIIChange{Most low-mass} red giants have $L < 3,000\ \Lsun$, although a few TP-AGB stars are brighter.  Of course, actual galaxies are a combination of these two groups.}

\begin{figure*}
\centerline{\includegraphics[width=9cm]{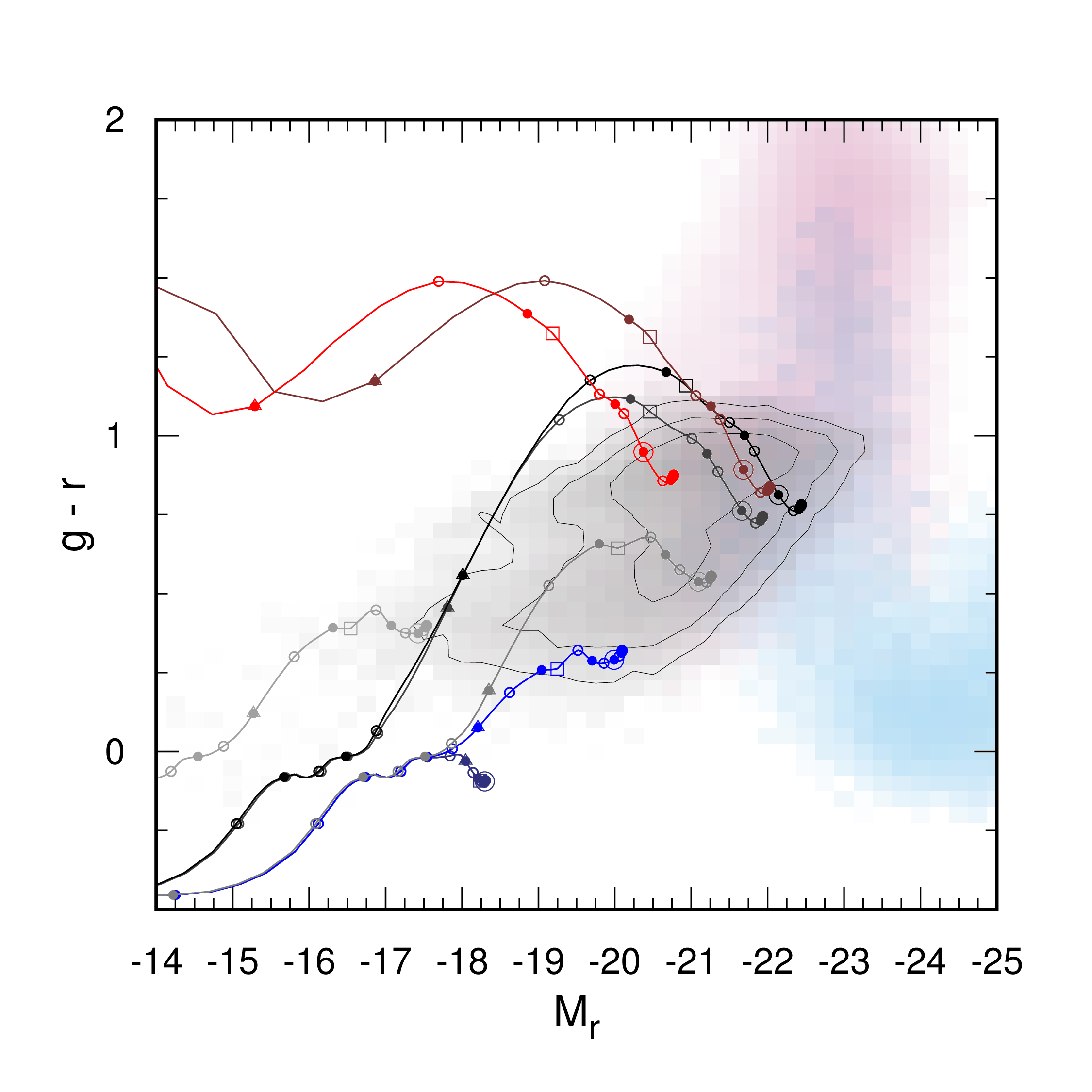}\includegraphics[width=9cm]{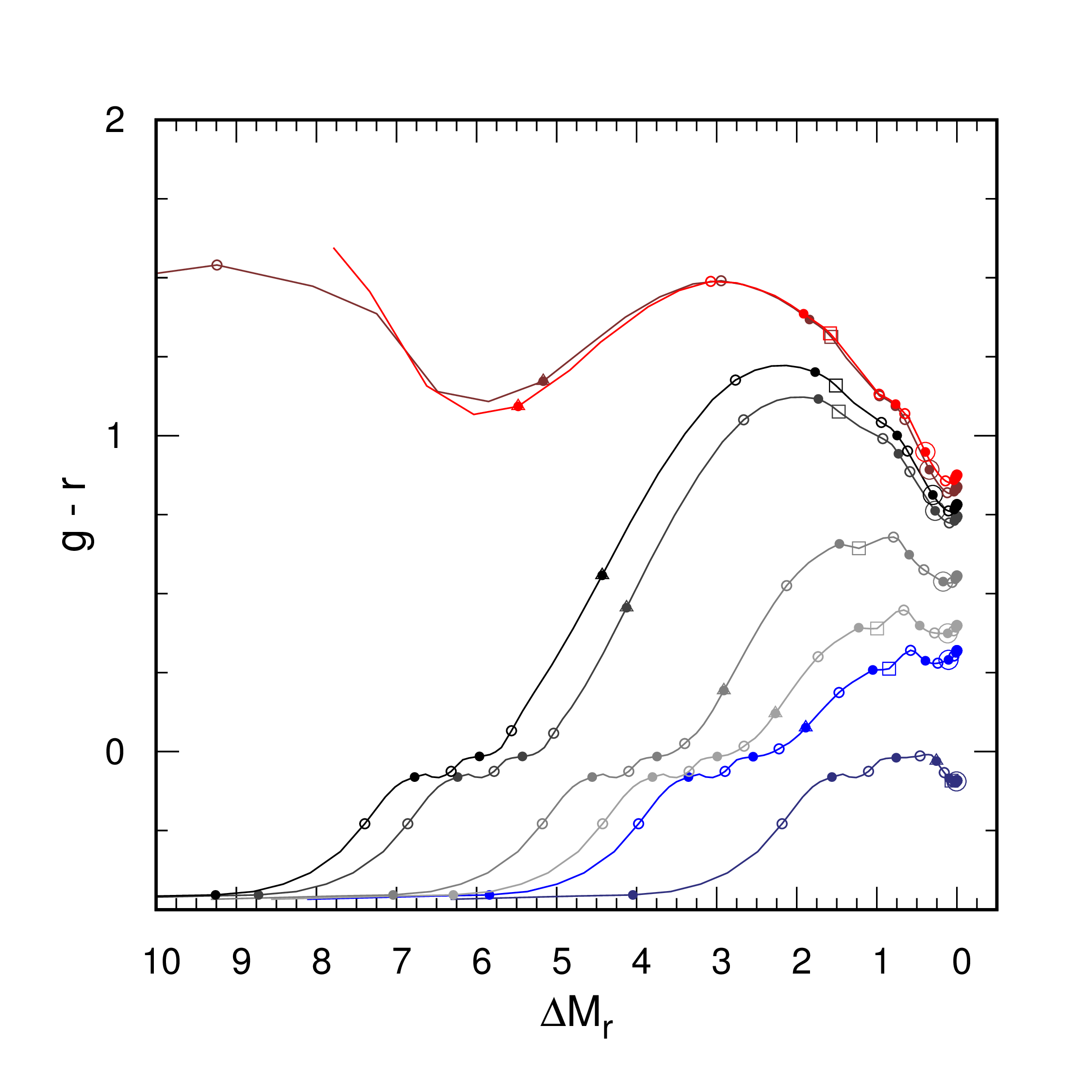}}
\figcaption{Color magnitude diagram for partially cloaked galaxies.  \vIIIChange{Unscreened galaxies start at the terminus of its track marked by nearly-overlapping solid points.  The effect of increasing $L_{\rm min}$ is to move the galaxy in color space along the track.} Tracks have the same colors \vIIIChange{for each SFH} as Figure~\ref{fig:MBol}.  A circle surrounds the points on the tracks with a threshold of $1\ \Lsun$, the triangle is around the point for a threshold luminosity of $1,000\ \Lsun$, and the squares are at the red clump luminosity ($10^{1.7}\ \Lsun$).  Small filled dots mark out powers of ten in $L_{\rm min}/\Lsun$, while small open dots mark out half-powers of ten.  \vIIIChange{On left, the absolute magnitudes of the model galaxies are shown; on right, the change in $r$-band magnitudes are plotted.} For comparison, a sample of SDSS objects are plotted \vIIIChange{on left}: \vIIChange{$z < 0.1$ galaxies in \vIIIChange{grey} shading and contours, $z \ge 0.1$ galaxies in \vIIIChange{reddish-purple} shading, and QSOs in \vIIIChange{sky blue} shading.}}
\label{fig:SunscreenCMD}
\end{figure*}

\begin{figure*}
\centerline{\includegraphics[width=9cm]{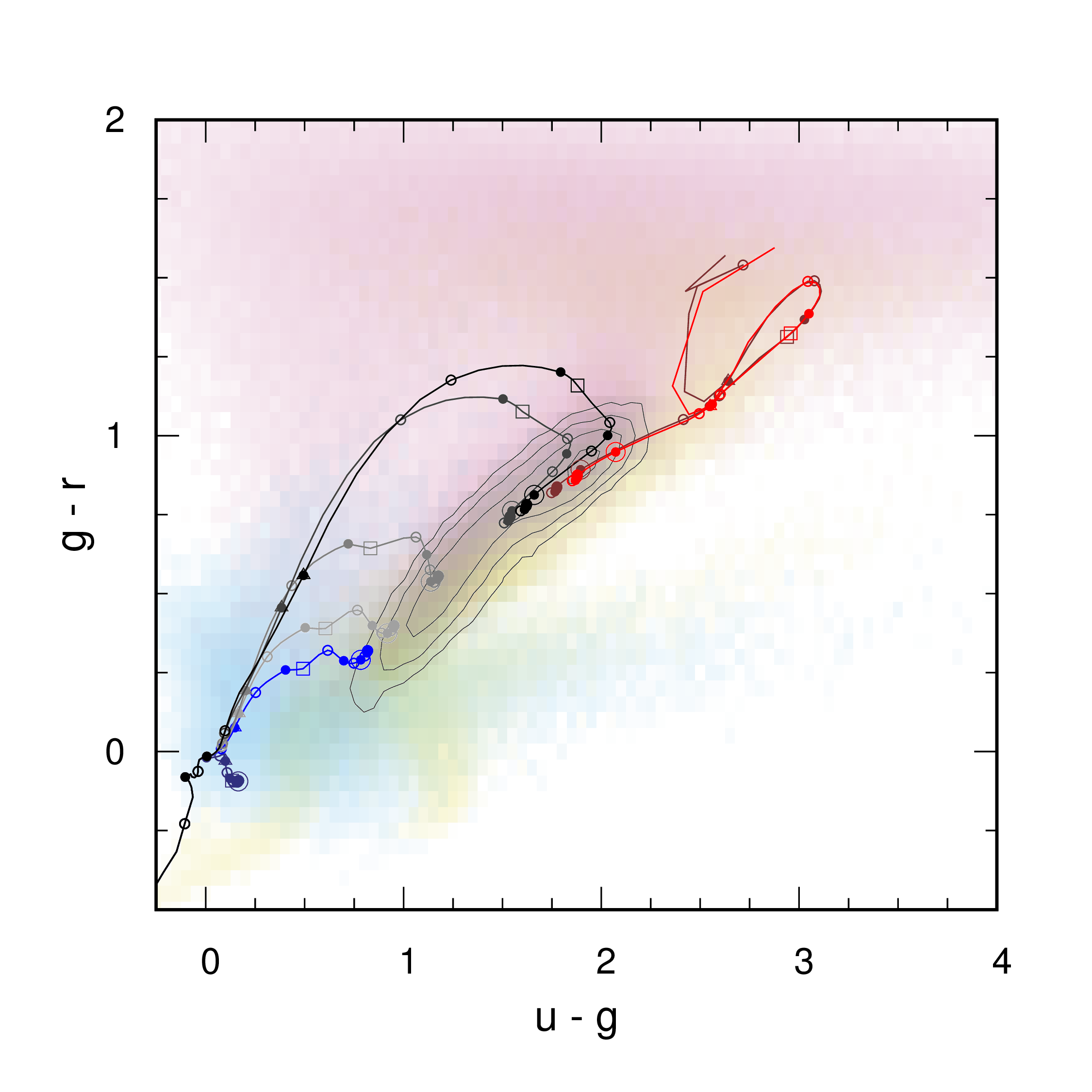}\includegraphics[width=9cm]{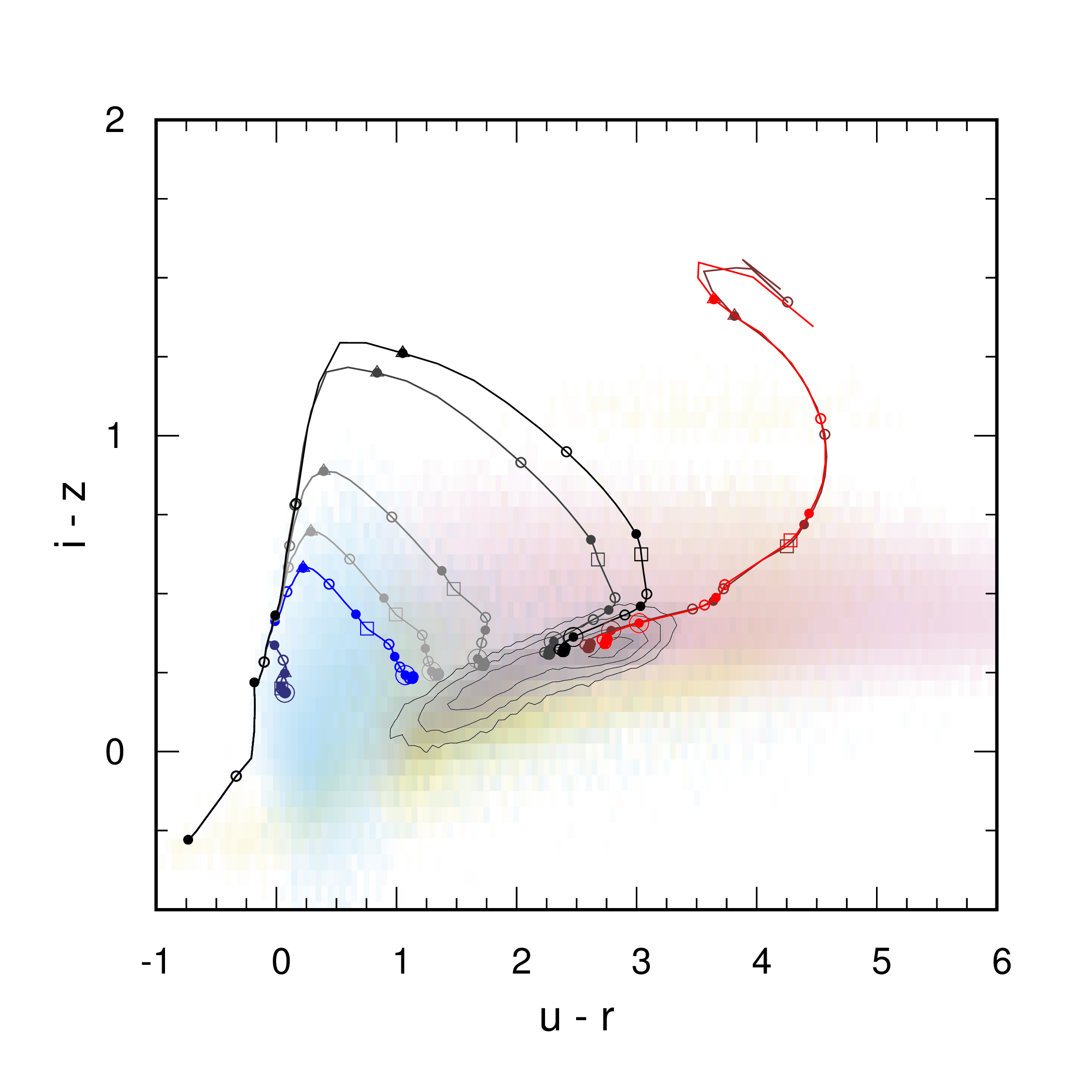}}
\centerline{\includegraphics[width=9cm]{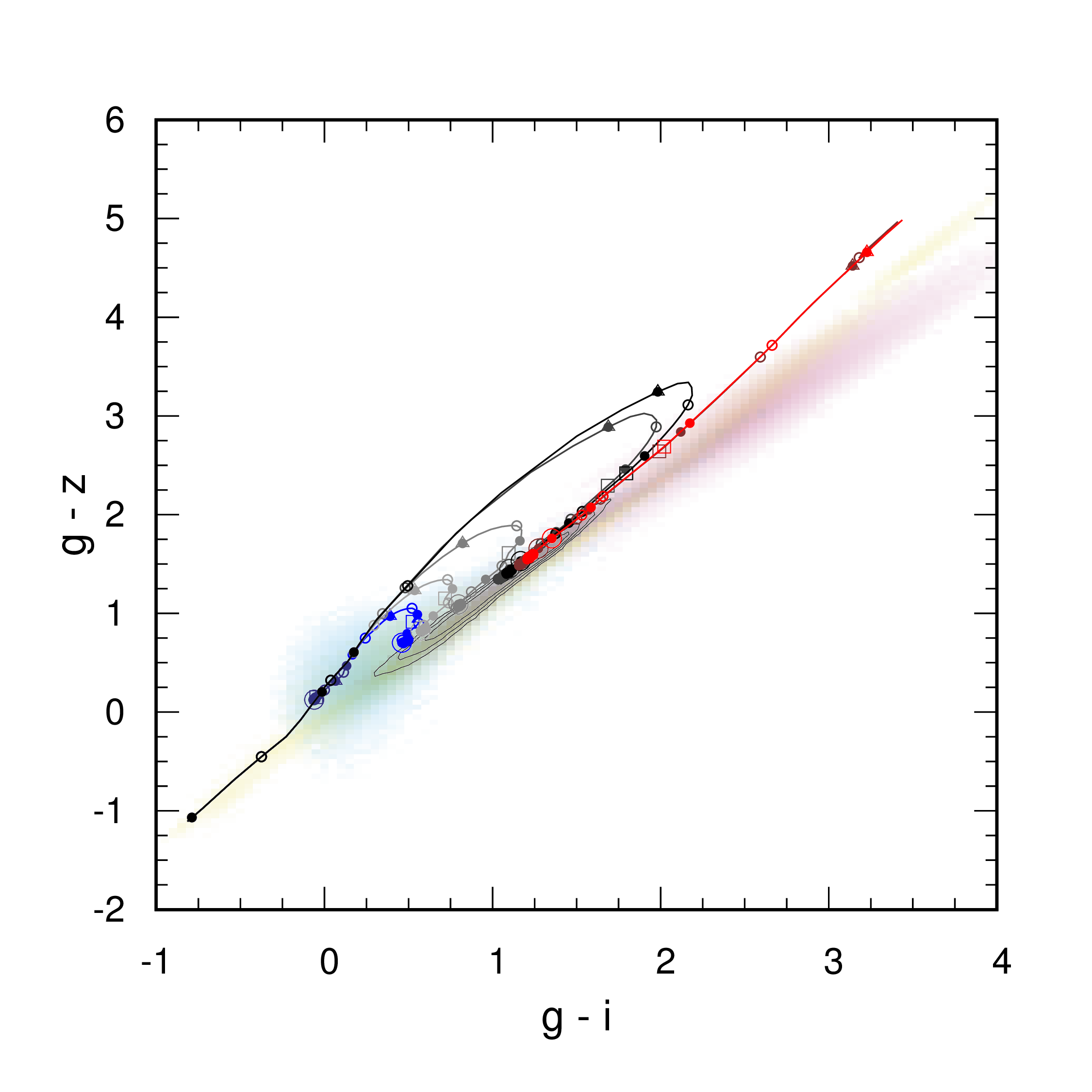}\includegraphics[width=9cm]{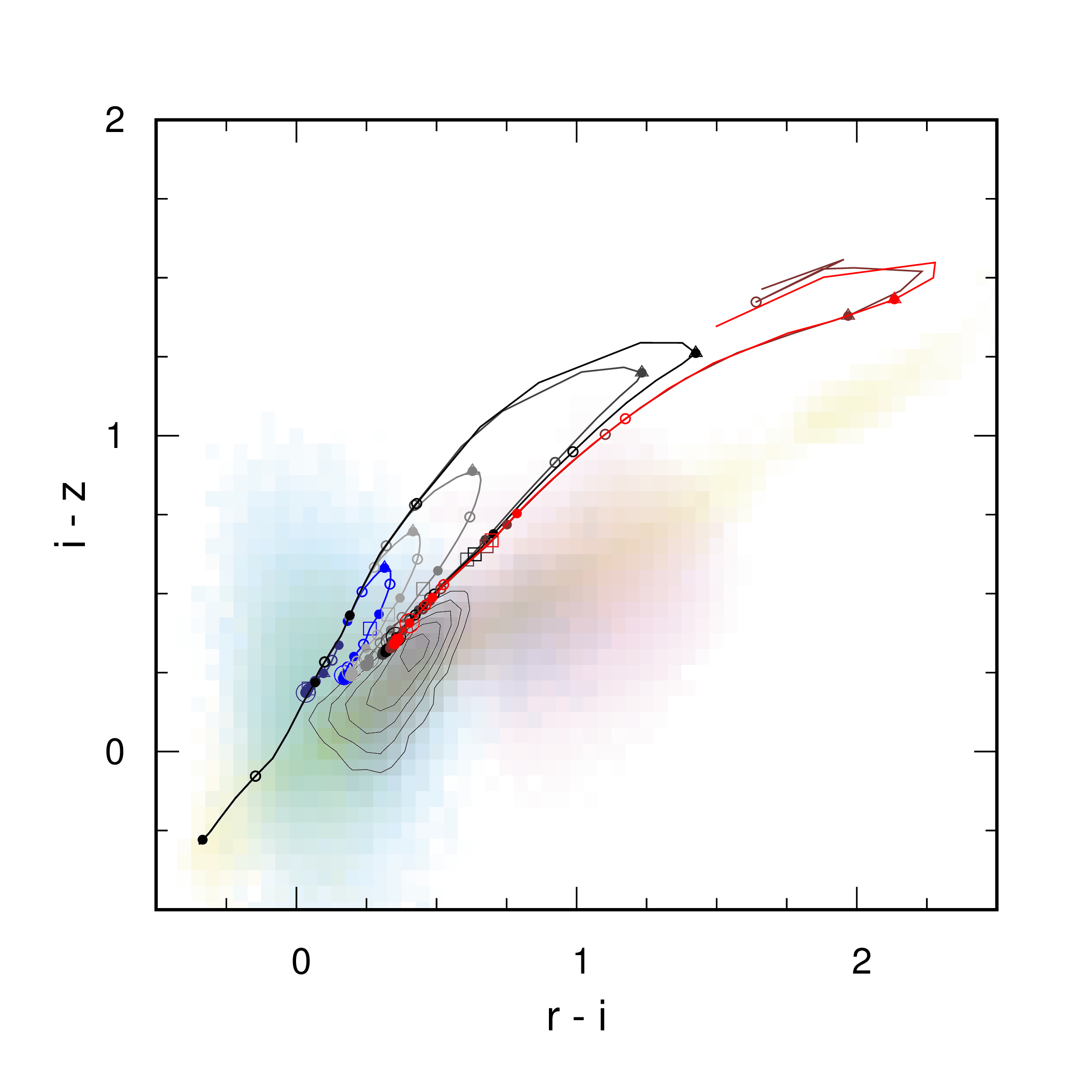}}
\figcaption{Effects on the colors of partially cloaked galaxies in color-color diagrams.  The yellow \vIIChange{shading represents stars selected from SDSS}; all other notations are the same as in Figures~\ref{fig:MBol}~and~\ref{fig:SunscreenCMD}.  \vIIIChange{Tracks have the same colors for each SFH as Figure~\ref{fig:MBol}.}\label{fig:SunscreenColors}}
\end{figure*}

\vIIChange{Ultimately, when searching for these galaxies in a photometric survey, we are interested in how partial cloaking moves a galaxy on color-magnitude (CMD) and color-color diagrams.  I show a visible-light CMD (Figure~\ref{fig:SunscreenCMD}) and three color-color diagrams (Figure~\ref{fig:SunscreenColors})that demonstrate the qualitative range of behaviors partial cloaking caused as compared to natural galaxies.  To show how the screened galaxies stand out from natural galaxies, I downloaded a sample of 500,000 objects in the SDSS spectroscopic database for SDSS Data Release 14, which are broadly classed into galaxies, quasars, and stars.}\footnote{\vIIChange{I performed a query using} the SkyServer, http://cas.sdss.org/dr14/en/tools/search/SQS.aspx\vIIChange{, retrieving the top 500,000 objects found in the spectroscopic database with clean redshifts and clean photometry.  I separated the galaxies into those with $z < 0.1$ and those with $z \ge 0.1$, but otherwise did not perform any distance selection.  Nor did I select on luminosity or object class.  The sample should be a uniformly random selection among objects with spectra in SDSS DR 14 that pass the simple data quality cuts.}}  Natural galaxies at $z = 0$ clearly separate into the red sequence and blue cloud in optical diagrams \citep[][ among many others]{Strateva01,Baldry04}.  \vIIChange{In infrared color-color diagrams (represented here by $(g-i)\endash (g-z)$ and $(r-i)\endash(i-z)$), the natural $z = 0$ galaxy populations tend to be stretched into narrow lines, providing a relatively clean way to find anomalous galaxies.}

\vIIChange{A galaxy with a fixed stellar population forms a track on these diagrams as $L_{\rm min}$ increases, as shown in the Figures.  The shapes of these tracks can also be understood from the relative contribution of dwarf and red giant stars to variation in $\alpha_x$.  \vIIIChange{Galaxies with $L_{\rm min} \approx 0$ start with positions on these diagrams near observed galaxies with similar SFHs: among the blue cloud for continuing-SF galaxies (although the 100 Myr constant SFR model is bluer than most observed galaxies; dark blue); in the green valley for declining-SF galaxies; and in the red sequence for quenched galaxies.  My model galaxies are somewhat offset from the observed galaxies by around 0.1--0.2 magnitudes.  The offset seems to be especially noticeable in diagrams with infrared colors, and is largest for continuing-SF galaxies.}  Galaxies do not move appreciably until $L_{\rm min} \ga 1\ \Lsun$ because the faintness of these stars cannot compensate for their larger number \vIIIChange{(indicated by the pile up of points near the start of the track)}.}

\vIIChange{As brighter stars are cloaked, though, the tracks leave the red sequence and blue cloud where most $z = 0$ galaxies reside.  In the $r \endash (g-r)$ CMD (Figure~\ref{fig:SunscreenCMD}), quenched and declining-SF galaxies veer towards red colors because their luminosity is dominated by red giants\vIIIChange{.  C}ontinuing-SF galaxies drift towards blue colors as \vIIIChange{red giants and low to intermediate-mass MS stars are cloaked, while O and B dwarfs remain unscreened.}  Ultimately, all galaxies \vIIIChange{with ongoing star formation} start becoming bluer when all the red giants are cloaked\vIIIChange{.}  Galaxies with recent star-formation end up along a terminal locus in color-color diagrams (Figure~\ref{fig:SunscreenColors}), with blue colors reminiscent of quasars.  

Observationally, the effects of increasing $L_{\rm min}$ can be summed up as: \emph{red galaxies get redder (especially in red bands), and blue galaxies get bluer (especially in blue bands).}  In infrared colors, continuing SF galaxies also \vIIIChange{initially} get redder because the post-MS stars dominate in these bands\vIIIChange{ -- only when all red giants are screened do these galaxies become bluer again.}.  The tracks from continuing-SF and quenched galaxies leave the main concentrations of galaxies on the color-color diagrams, implying partially cloaked galaxies with $L_{\rm min} \ga 1\ \Lsun$ can be selected with a color cut.  Quenched galaxies have particularly large excursions, becoming over a magnitude redder or more than the red sequence once the red clump is screened (squares on Figure).  The excursions are also significant in the infrared color-color diagrams simply because \vIIIChange{stars and most} $z = 0$ galaxies are so concentrated in these diagrams.  \vIIIChange{Deviations from the stellar line on $(g-i)\endash (g-z)$ plot arise from the changing balance of the unscreened stellar populations that produce optical light and those producing NIR light.}

Despite these variations, the altered galaxies share color-color space with other source populations.  The extra-blue galaxies tend to have similar colors as quasars, while the extra-red galaxies tend to have similar colors as high-redshift galaxies.  Other criteria must therefore be implemented to select out these potential false positives: quasars are bright and compact, while partially cloaked galaxies would be dim (as seen in the CMD) and are extended; spectroscopically measured redshifts can weed out red, high-$z$ galaxies.}

\subsection{\vIIChange{Variants}}
\label{sec:Variants}
\vIIChange{With the code, it is possible to investigate how galaxies with a variety of different stellar populations are affected by partial cloaking.  Among the other effects I considered are metallicity, galaxy age, IMF, the use of $(L/M)$ instead of $L$ as a selection cut, and the effects of cloaking the brightest stars only.  These variant tracks are plotted for a continuing-SF galaxy and a quenched SFH galaxy in Figure~\ref{fig:CCDVariants}.

Some of these parameters have little qualitative effect on the tracks of galaxies on color-color and color-magnitude diagrams.  For the reasons described in Section~\ref{sec:Spectra}, using luminosity-to-mass ratio as a cut (dotted line) instead of luminosity does not substantially change the tracks.  Changing the population age -- using the rest-frame colors of a galaxy as it was billions of years ago instead of today -- conceivably could matter, since the main sequence turn-off location is shifted to more massive, bluer stars.  For continuing-SF galaxies, changing the population age to \vIIIChange{6\ \Gyr} ($10^{9.8}\ \yr$) or \vIIIChange{3\ \Gyr} ($10^{9.5}\ \yr$) has little effect (blue lines).  Some currently quenched galaxies were in fact still forming stars at these times, and $z = 0$ declining-SF galaxies had more vigorous star formation back then, so these galaxies would have behaved as though their luminosity-weighted ages were smaller.  Despite the larger main sequence turn-off mass, a galaxy with a pure burst SFH at $z \approx 1$ would behave much like one at present day with regards to partial cloaking.  \vIIIChange{Another change with an effect similar to altering the SFH is} using a bottom-heavy IMF\vIIIChange{. It} favors \vIIIChange{old, low mass} red giants at the expense of young blue dwarfs, making the galaxy behave as if its SFH were more quenched (red line).

It is disputed whether or not a galaxy's metallicity affects its chances of hosting intelligent life \citep{Gonzalez01}, but starfaring ETIs might spread to a low metallicity galaxy and engineer it.  While metallicity does not generally affect the shapes of tracks on these diagrams, it does control how far they venture from the bulk of $z = 0$ galaxies.  Low metallicity galaxies, with $Z = 0.1\ \Zsun$, have brighter and bluer stars, with $\alpha_x$ declining more slowly and more synchronously in optical bands (Figure~\ref{fig:MagnitudeFunctions}).  Thus, these galaxies simply do not wander much on color-color diagrams as $L_{\rm min}$ increases, and their tracks generally do not leave the bulk of $z = 0$ galaxies (solid black line).  The fluxes from high metallicity galaxies (here, $Z = 2\ \Zsun$), in contrast, are more sensitive to partial cloaking, as demonstrated by their wider excursions on the color-color plots (dash-dotted black line).

\begin{figure*}
\centerline{\includegraphics[width=9cm]{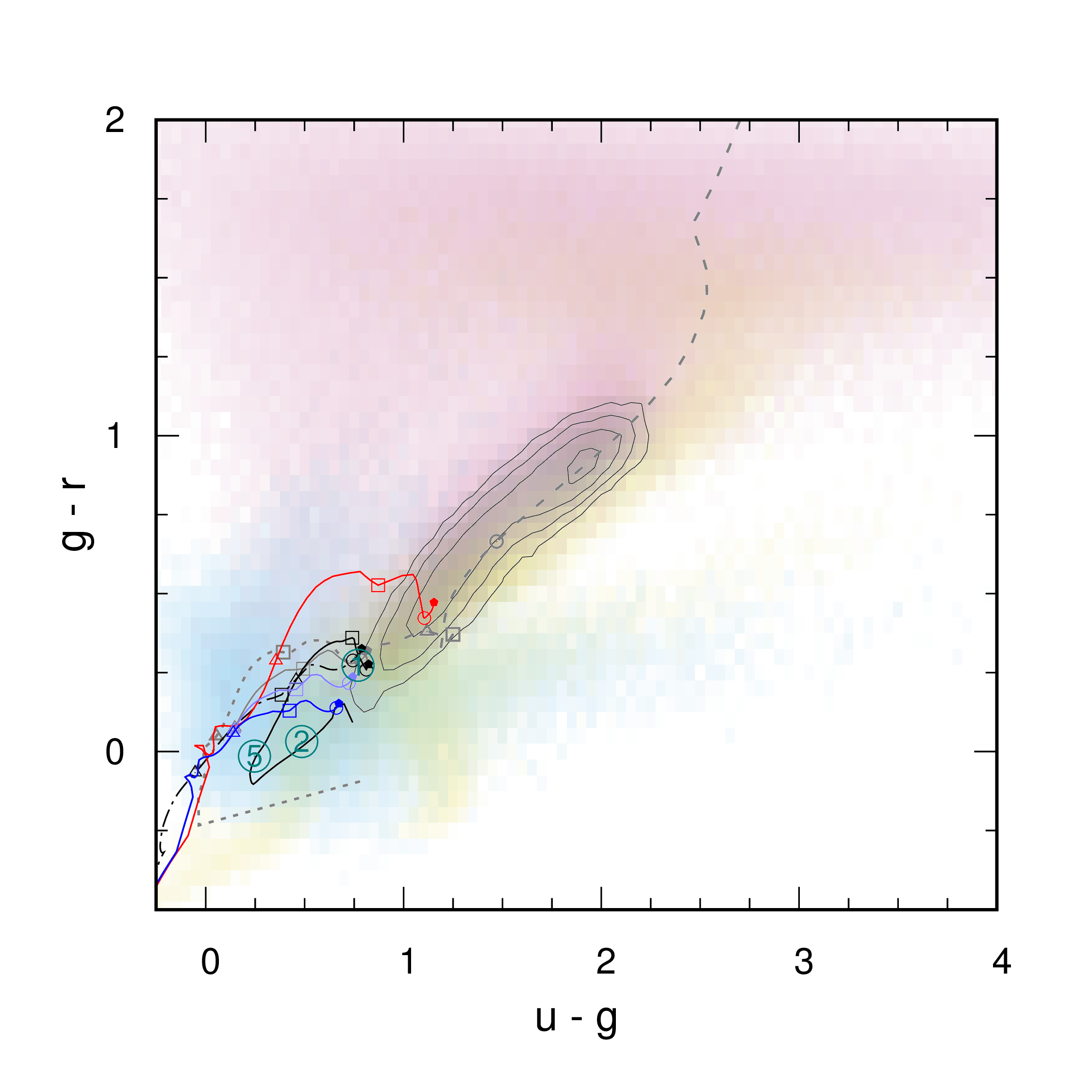}\includegraphics[width=9cm]{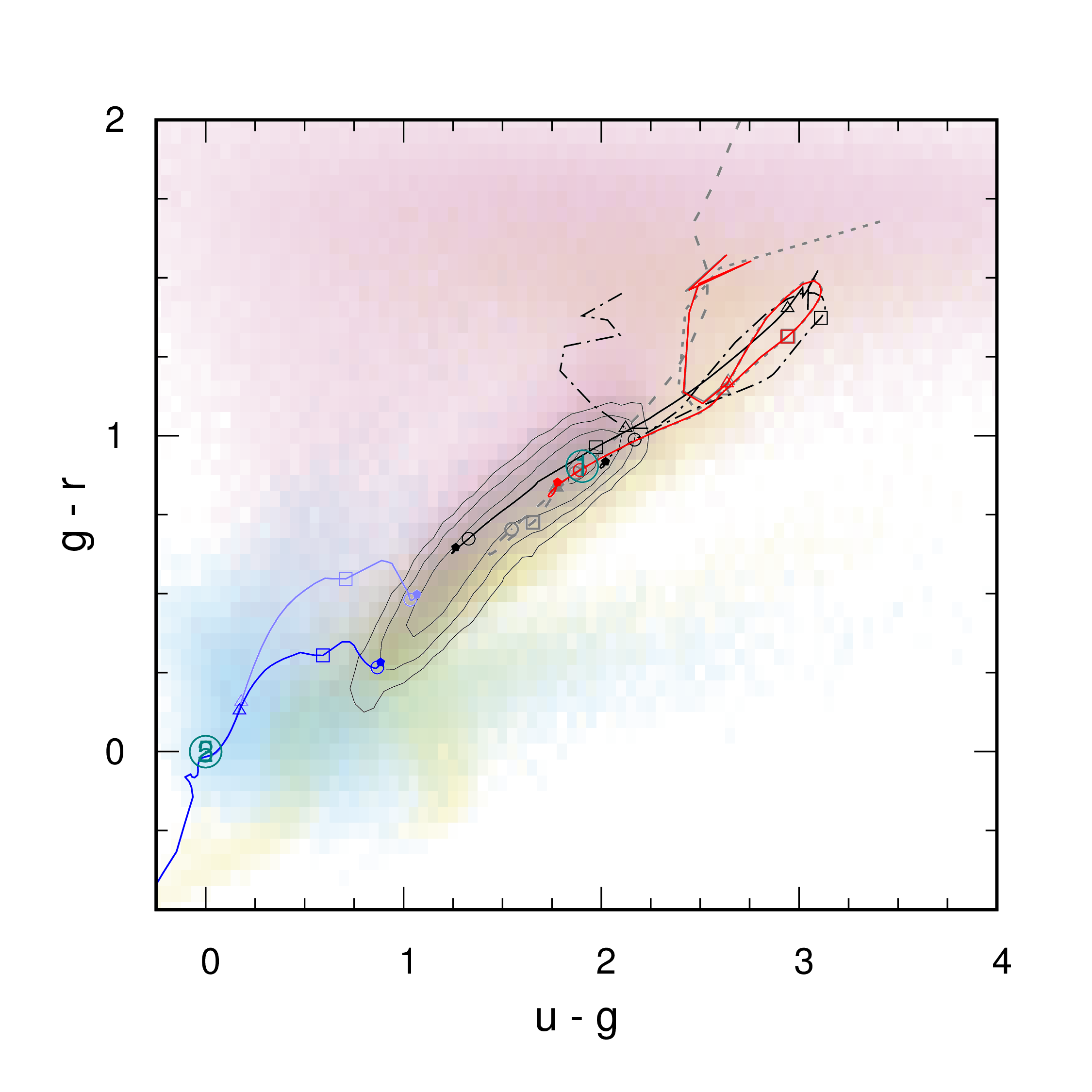}}
\centerline{\includegraphics[width=9cm]{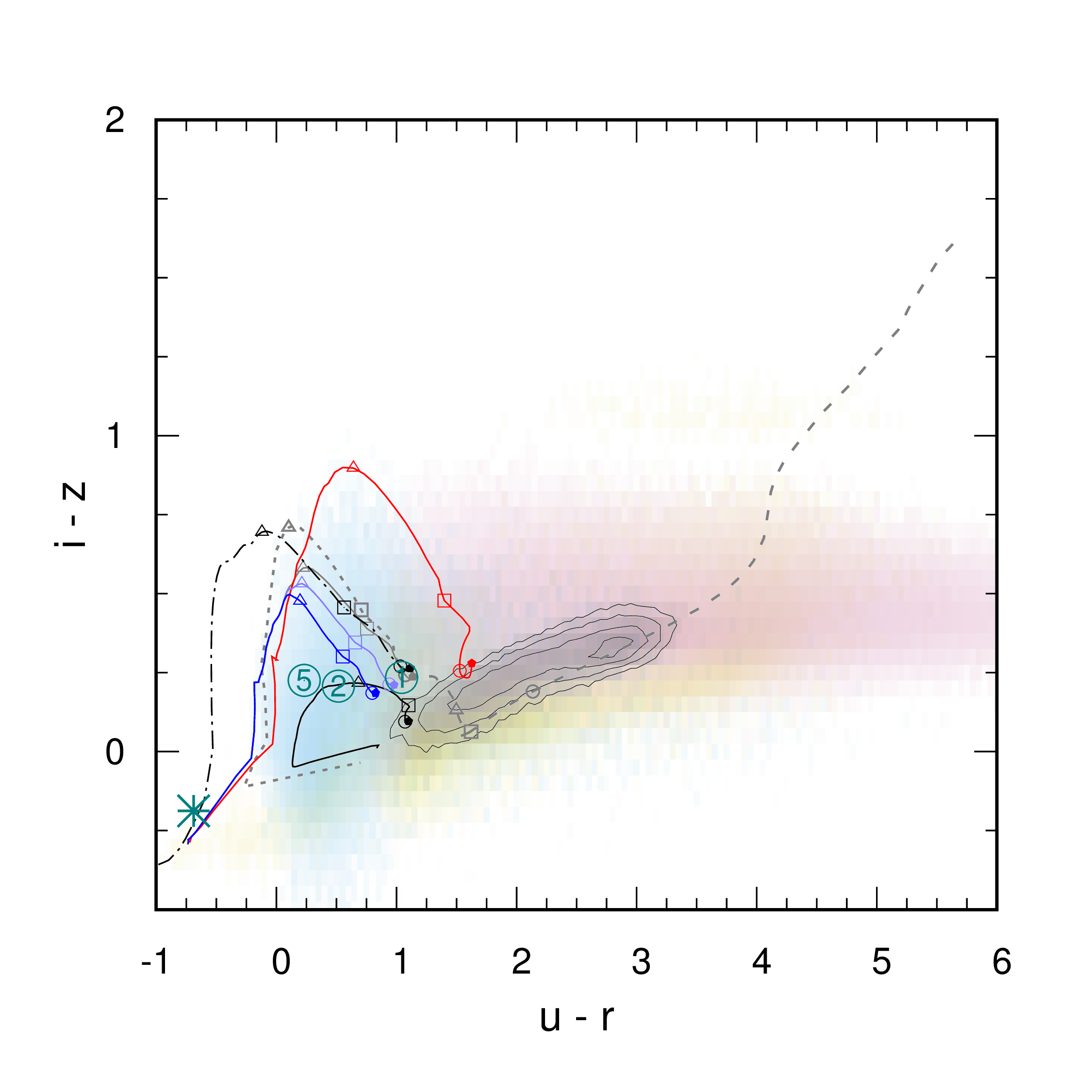}\includegraphics[width=9cm]{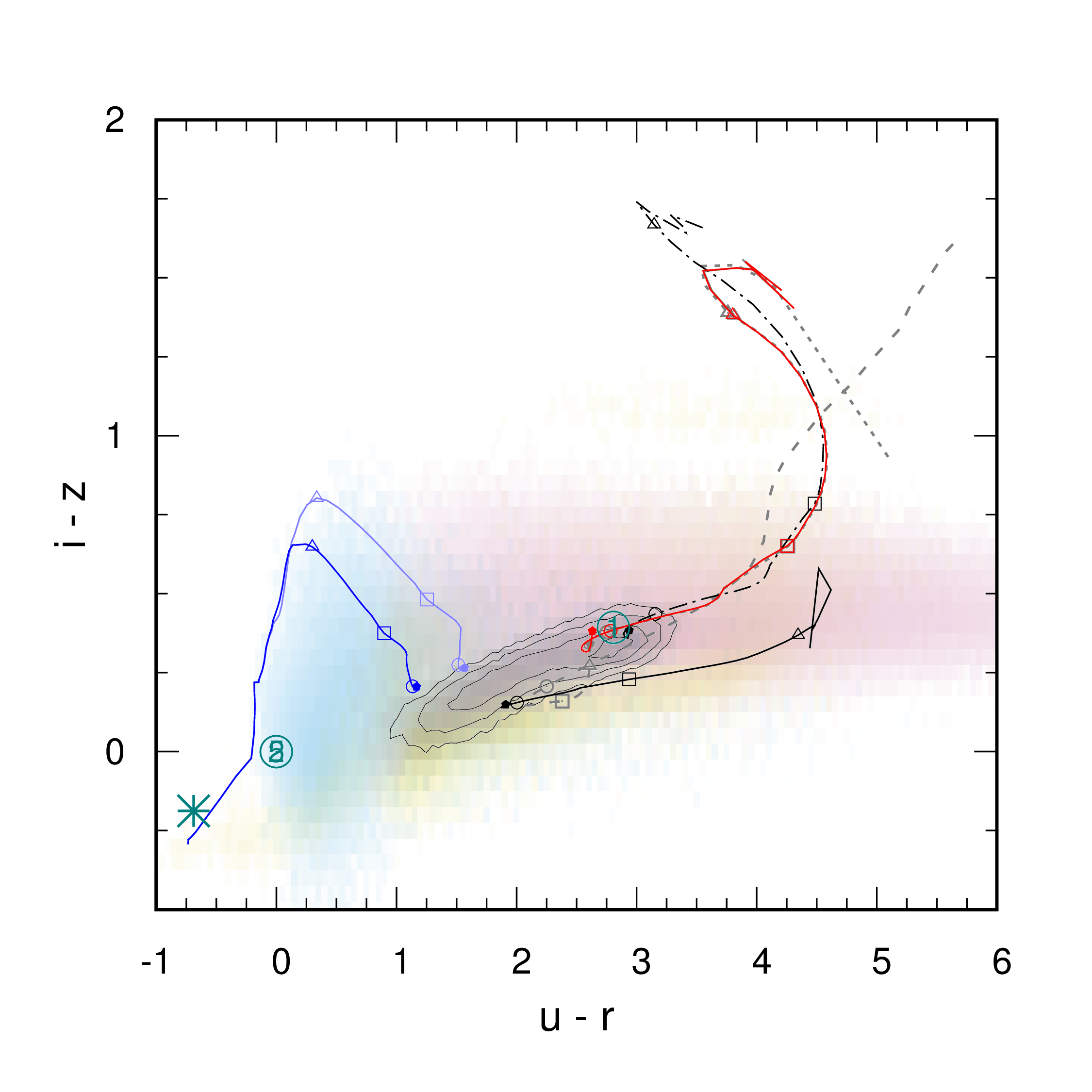}}
\figcaption{\vIIChange{The tracks of partially cloaked galaxies on color-color diagrams when I vary the assumptions.}  Representatives of \vIIIChange{old} continuing-SF (left) and quenched (right) \vIIIChange{galaxies} are shown.  Plotted variations are: fiducial (grey solid), brightest stars screened (dashed), $(L/M)_{\rm min}$ as the threshold for screening (dotted), low metallicity (black solid), high metallicity (black dot-dashed), a bottom-heavy IMF (red solid), and galaxy ages of \vIIIChange{$6\ \Gyr$} (pale blue) and \vIIIChange{$3\ \Gyr$} (bright blue).  The small, filled pentagons mark the unscreened start of the track.  Also shown are unscreened colors and magnitudes for IMFs with masses below $1$, $2$, and $5\ \Msun$ (marked by the turquoise circled numbers) and a young starburst (turquoise star).}  All other notations are the same as in Figure\vIIChange{s}~\vIIIChange{\ref{fig:MBol}, \ref{fig:SunscreenCMD}, and~\ref{fig:SunscreenColors}}.\label{fig:CCDVariants}}
\end{figure*}

One variation does display qualitatively different behavior: cloaking stars brighter than a threshold luminosity $L_{\rm max}$ (dashed grey line).  There could be various motivations for preferentially shrouding bright stars.  These stars have a high luminosity-to-mass ratio and are therefore especially suitable for ``starlifting'', in which a star's own luminosity is harnessed to unbind its envelope and mine it for materials \citep{Criswell85}.  In general, if aliens want a site where vast amounts of power are concentrated on sub-parsec scales, if AGNs or compact objects are unsuitable sites, and if they are unable or unwilling to beam power across interstellar distances, they would preferentially enclose brighter stars.  One possible application is to transmit vast amounts of information across interstellar distances \citep{Kardashev64}.

Interestingly, screening the very brightest stars in a stellar population doesn't affect its simulated photometry much.  \vIIIChange{Stars} with $L \gg 10^3\ \Lsun$ are characteristic of young stellar populations and are present mainly in galaxies with recent star formation.  Even in the \vIIIChange{old constant-SFR} galaxy \vIIIChange{model}, they account for only 60\% of the bolometric luminosity (Figure~\ref{fig:MBol}), so decreasing $L_{\rm max}$ from $\infty$ to $1,000\ \Lsun$ increases the bolometric magnitude by at most $1$.  As seen in Figure~\ref{fig:MagnitudeFunctions}, these stars emit the majority of the light in ultraviolet but a minority of visible and infrared light.  Thus, \vIIIChange{unless a galaxy has had a recent or ongoing starburst, the changes in its colors induced by $L_{\rm max} \gg 1,000\ \Lsun$ screening do not stand out against the natural variance in galaxy colors.}

\vIIChange{Even as $L_{\rm max}$ decreases to $1\ \Lsun$, the tracks do not verge into unnatural regions of color-color space.  For these luminosity thresholds, the brightest and reddest post-MS stars are screened, as well as more and more relatively blue dwarfs.  The remaining stellar population, forming the lower mass main sequence and the lower luminosity red giant branch, combine together to maintain integrated colors that are not wildly divergent from natural galaxies.  There is a reversal of the normal rule-of-thumb for color deviations in this $L_{\rm max}$ range -- red galaxies get slightly bluer and blue galaxies get slightly redder, especially in redder bands -- but the effect is comparatively minor.  Only when $L_{\rm max} < 1\ \Lsun$, leaving only cool dwarfs unscreened, does the integrated stellar light become unnaturally red for all galaxies.}

If the radiation is processed into infrared waste heat, however, it would be a notable signature of these galaxies.  In the AGENT formulation, $\alpha$ passes the \GHAT~threshold of $0.25$ for all types of galaxies when $L_{\rm max} \ga 300\ \Lsun$; for \vIIIChange{old} \vIIChange{continuing-SF} galaxies, the threshold is $\sim 10^5\ \Lsun$ (Figure~\ref{fig:MBol}).  Most of the starlight is reprocessed when $L_{\rm max}$ is $\sim 30\ \Lsun$ in \vIIChange{quenched} and $3,000\ \Lsun$ in \vIIIChange{old} \vIIChange{continuing-SF} galaxies.  The waste heat would make these galaxies look like highly obscured active galactic nuclei (AGNs), or starburst galaxies if the waste heat was cold enough -- except that the emission comes from an unusually extended region ($\sim 10\ \kpc$) rather than a compact core.  If instead the waste heat is emitted with habitable temperatures, the galaxies would appear as diffuse sources emitting only MIR and no FIR, a very unusual combination.

\subsection{\vIIChange{Degeneracies with Alternate IMFs and SFHs?}}
\vIIChange{Some of the unnatural alterations to galaxy photometry described here are degenerate with unusual IMFs.  Because the main sequence forms a monotonic relation with mass, screening parts of the main sequence is equivalent to suppressing star formation in that mass range.  A galaxy with $L_{\rm min} \la 1\ \Lsun$ looks much the same in the optical or UV as one with no low mass dwarfs; one with $L_{\rm max} \gg 1,000\ \Lsun$ looks much like a galaxy where massive stars never formed or have already died out.  

In Figure~\ref{fig:CCDVariants} the circled numbers represent simulated photometry of galaxies with all star formation suppressed below $1$, $2$, and $5\ \Msun$, corresponding to main-sequence turnoff ages of roughly 10, 1, and 0.1 Gyr.  The bluer optical colors of these bottom-light IMF galaxies are comparable to those of partially cloaked continuing-SF galaxies.  A very young starburst can also have extreme blue colors, as shown by the star point in the figures denoting a starburst that is only \vIIIChange{1} Myr old.  However, the alternate IMFs suppress the red giant population of these galaxies -- there are no redward excursions like those seen for partially cloaked quenched galaxies, particularly in infrared colors.  Unnaturally red galaxies would require a narrow peak in the IMF around $1\ \Msun$.

Any level of cloaking can be emulated by adjusting the joint IMF-SFH distribution arbitrarily.  Stars of a given age and mass tend to all have nearly the same luminosity: that's what allows us to use isochrones in the first place.  So by adjusting the joint age-mass distribution for all stars with a given luminosity, any degree of screening can be accommodated.  In that case, only non-optical methods could provide evidence for screening specifically, like dynamical estimates of mass to light or waste heat searches.  When $L_{\rm min} \approx 10 \endash 1,000\ \Lsun$, though, the joint distributions are likely to be contrived because of the importance of red giants in this luminosity range.  As noted before, red giants tend to have nearly the same mass because the post-MS phase of stellar evolution is relatively short.  

More generally, it is an empirical fact that most galaxies don't have the extreme colors of the more divergent partially cloaked galaxies.  Any galaxy with such a strange IMF and SFH would stand out in a photometric survey.  While we could not be entirely sure that partial cloaking is responsible, it would be a clear sign that something bizarre is going on and that it merits a closer look.}

\section{Conclusions}
\label{sec:Conclusion}
Dyson spheres are hard to build, and they are especially hard to build around bright stars.  Yet bright stars are the source of most of the bolometric luminosity in a non-active galaxy.  A megastructure-oriented society that would otherwise be Type III may therefore fail to process most of a galaxy's starlight if it is based on the classic Dyson sphere concept.  Instead, it might only enclose stars below a certain luminosity threshold $L_{\rm min}$ determined by practical constraints.  These societies could be easily missed by previous searches for Type III societies, those searching for vast amounts of waste heat or profound optical dimming.  Alternatively, a nearly Type III society might only enclose the brightest stars.

I have developed a stellar population synthesis code that allows me to compute the spectrum of a galaxy that appears to be missing its brightest or dimmest stars.  From the output spectra, I can then calculate the observed magnitude of the partially cloaked galaxy to search for signatures that could be sought in photometric surveys.  The advantage of using photometric surveys is that they catalog enormous numbers of galaxies.  In addition, by searching for magnitude-color signatures of missing stars, we do not need to know the form the waste heat takes or even whether there is waste heat.  I calculated the spectra for a variety of SFHs (Figure~\ref{fig:SFH}).

My general result is that in visible light colors, when $1\ \Lsun \la L_{\rm min} \la 1,000\ \Lsun$, \emph{red galaxies get redder while blue galaxies get bluer}.  This can be understood from the fact that bright stars tend to be either young, massive, blue stars, or they are red giants.  The former are responsible for one peak in a galaxy's spectrum at ultraviolet energies, while the latter are responsible for another peak in visible to NIR.  The blue stars are the brightest (up to $10^6\ \Lsun$), while low mass red giants tend to have a peak luminosity at $\sim 1,000\ \Lsun$.  Thus, \vIIChange{continuing-SF} galaxies get bluer in the optical because the red giants go ``missing'', leaving the brighter blue stars.  \vIIChange{Quenched} galaxies get redder \vIIIChange{after negligible evolution when $L_{\rm min} \ll 1\ \Lsun$} because the stars that \vIIIChange{next} go ``missing'' \vIIIChange{as $L_{\rm min}$ increases} are the relatively \vIIIChange{hot} \vIIIChange{G} dwarfs\vIIIChange{, the} subgiants\vIIIChange{,} and \vIIIChange{then the relatively hot red giants at the base of the RGB}.  In near-infrared, all galaxies get redder, since starlight at these wavelengths is dominated by red giants, and to a lesser extent, \vIIIChange{low mass} main sequence stars\vIIIChange{, just like in quenched galaxies}.

When $L_{\rm min} \la 1\ \Lsun$, there are only small effects on the colors of a galaxy.  The deviations are greatest for \vIIChange{quenched} galaxies, with no luminous young stars to smother the signal, and in the near-infrared for the same reason.

I also tested the effects of variant mode populations with different parameters.  \vIIChange{Most importantly, galaxies where only the brightest stars are cloaked behave qualitatively different.  They have colors typical of $z = 0$ galaxies for luminosity cuts above $1\ \Msun$ and then become unnaturally red when only faint \vIIIChange{dwarf} stars remain.}

Assuming megastructure-building, galaxy-spanning societies exist within the observable Universe, should we expect to find partially cloaked galaxies with these spectra and colors?  In this paper, I have assumed that there is one homogeneous threshold luminosity that applies throughout a galaxy.  That may not be realistic, since different regions of a galaxy are so distant from each other that they may not be able to coordinate a unified program of megastructure building.  In addition, different regions of a galaxy that we observe simultaneously are separated \emph{temporally} from each other, and the agenda of a galactic society could change with time \citep[as in][]{Hart75,Wright14-SF}.  Instead, it's possible that there are galaxies with patchworks of adjusted stellar populations.  They would appear like a mosaic of tiles with diverse colors and brightnesses.  They could be found by looking at resolved galaxies, but it would take a more detailed analysis.  Over time, these stellar populations would mix together and dilute any signature, unless the stars are being actively steered (as described in \citealt{Badescu06}).

Galaxies in the Local Group are close enough that their color-magnitude diagrams can be constructed through direct observation.  It should be possible to search among the Group for parts of galaxies that are ``missing'' stars below (or above) a certain luminosity.  There would have to be a way to distinguish artificial engineering from IMF variations.  

It's also possible that a galaxy-spanning society would not merely cloak stars, but shape their formation and evolution more directly.  This could lead to galaxies with ``impossible'' types, or stellar phenomena appearing completely out of proportion to their natural values.  We could search for galaxies with a large number of hypermassive stars, for example, much greater in mass than the $\sim 100 \endash 200\ \Msun$ stars we know of.  The stellar population of a galaxy might somehow have an unnaturally high metallicity, moving the galaxy off the mass-metallicity relation.  Rampant starlifting might lead to a galaxy with most of its brighter stars in a planetary nebula phase \citep{Lacki16-K3}.  Artificial stars might be created, with unnatural properties, designed to create astronomical amounts of metals.  

There are widescale interventions that we could look for, even if no stars with individually unnatural properties are fabricated.  The IMF of a galaxy may be adjusted in unusual ways, producing only stars of a certain mass for millions of years.  The SFH may also be adjusted in unnatural ways; for example, a \vIIChange{main sequence \vIIIChange{galaxy}} might completely shut off star-formation for a couple of billion of years despite having a supply of gas, and then resume it at its natural high rate for no apparent reason.  Either of these interventions might be detectable for billions of years through a detailed analysis of a galaxy's stellar population.  

\acknowledgements
I wish to acknowledge the support of the Breakthrough Listen program.  Funding for \emph{Breakthrough Listen} research is sponsored by the Breakthrough Prize Foundation\footnote{https://breakthroughprize.org/}.

In addition, I acknowledge the use of NASA's Astrophysics Data System and arXiv.  I also thank the referee for their thorough review of this paper.

The plots use data from the Sloan Digital Sky Survey.  Funding for the Sloan Digital Sky Survey IV has been provided by the Alfred P. Sloan Foundation, the U.S. Department of Energy Office of Science, and the Participating Institutions. SDSS acknowledges support and resources from the Center for High-Performance Computing at the University of Utah. The SDSS web site is www.sdss.org.  SDSS is managed by the Astrophysical Research Consortium for the Participating Institutions of the SDSS Collaboration including the Brazilian Participation Group, the Carnegie Institution for Science, Carnegie Mellon University, the Chilean Participation Group, the French Participation Group, Harvard-Smithsonian Center for Astrophysics, Instituto de Astrofísica de Canarias, The Johns Hopkins University, Kavli Institute for the Physics and Mathematics of the Universe (IPMU) / University of Tokyo, the Korean Participation Group, Lawrence Berkeley National Laboratory, Leibniz Institut für Astrophysik Potsdam (AIP), Max-Planck-Institut für Astronomie (MPIA Heidelberg), Max-Planck-Institut für Astrophysik (MPA Garching), Max-Planck-Institut für Extraterrestrische Physik (MPE), National Astronomical Observatories of China, New Mexico State University, New York University, University of Notre Dame, Observatório Nacional / MCTI, The Ohio State University, Pennsylvania State University, Shanghai Astronomical Observatory, United Kingdom Participation Group, Universidad Nacional Autónoma de México, University of Arizona, University of Colorado Boulder, University of Oxford, University of Portsmouth, University of Utah, University of Virginia, University of Washington, University of Wisconsin, Vanderbilt University, and Yale University.


\end{document}